\RequirePackage{fix-cm}
\documentclass[smallextended]{svjour3}       % onecolumn (second format)
\smartqed  % flush right qed marks, e.g. at end of proof
\usepackage{graphicx}
\usepackage{amsmath}
\usepackage{enumitem,amssymb}
\usepackage[authoryear]{natbib}  % Added by LIG 2020.12.05

\def\Mdot{$\dot M$}

\journalname{Experimental Astronomy}
\begin{document}

\title{THEZA: TeraHertz Exploration and Zooming-in for Astrophysics
}
\subtitle{An ESA Voyage 2050 White Paper}


\author{Leonid~I.~Gurvits        	\and
            Zsolt~Paragi                	\and
            Viviana~Casasola       	\and 
            John~Conway 		\and
            Jordy~Davelaar		\and
            Heino~Falcke 		\and
            Rob~Fender 			\and
            S\'{a}ndor~Frey 		\and
            Christian~M.~Fromm 	\and
            Cristina~Garc\'{\i}a~Mir\'o \and
            Michael~A.~Garrett 	\and
            Marcello~Giroletti        	\and   
            Ciriaco~Goddi 		\and
            Jos\a'e-Luis~G\a'omez 	\and
            Jeffrey~van~der~Gucht \and
            Jos\a'e~Carlos~Guirado 	\and
            Zolt\a'an~Haiman 		\and
            Frank~Helmich 		\and
            Elizabeth~Humphreys 	\and
            Violette~Impellizzeri 	\and
            Michael~Kramer 		\and
            Michael~Lindqvist 		\and
            Hendrik~Linz 		\and
            Elisabetta Liuzzo 		\and
            Andrei~P.~Lobanov 	\and
            Yosuke~Mizuno 		\and
            Luciano~Rezzolla 		\and
            Freek~Roelofs 		\and
            Eduardo~Ros 		\and
            Kazi~L.\,J.~Rygl 		\and
            Tuomas~Savolainen 	\and
            Karl~Schuster 		\and
            Tiziana~Venturi 		\and
            Martina~C.~Wiedner 	\and
            J.\ Anton~Zensus
}


\institute{L.I.~Gurvits \at
              Joint Institute for VLBI ERIC, Dwingeloo, EU / The Netherlands \\
              and \\
              Department of Astrodynamics and Space Missions, Delft University of Technology, \\
              The Netherlands \\
              \email{lgurvits@jive.eu}  \\
              ORCID: 0000-0002-0694-2459
               \and
              Z.~Paragi \at
              Joint Institute for VLBI ERIC, Dwingeloo, EU / The Netherlands 
              \and
              V.~Casasola \at
              INAF Institute of Radio Astronomy, Bologna, Italy 
              \and
              J.~Conway \at 
              Onsala Space Observatory, Chalmers University, Sweden 
              \and
              J.~Davelaar \at 
              Radboud University, Nijmegen, The Netherlands 
              \and
              H.~Falcke \at 
              Radboud University, Nijmegen, The Netherlands 
              \and
              R.~Fender \at 
              Oxford University, UK 
              \and
              S.~Frey \at
              Konkoly Observatory, Res. Centre for Astronomy and Earth Sciences, Budapest, Hungary \\
              and \\
              Institute of Physics, ELTE E\a"otv\a"os Lor\a'and University, Budapest, Hungary 
              \and         
              Ch.M.~Fromm \at 
              Goethe-Universit\"at Frankfurt, Germany 
              \and
              C.~Garc\'{\i}a Mir\'o \at
              Joint Institute for VLBI ERIC, Madrid,  EU / Spain 
              \and          
              M.A.~Garrett \at 
              Jodrell Bank Centre for Astrophysics, The University of Manchester, UK  \\
              and \\
              Leiden Observatory, Leiden University, The Netherlands \\
              ORCID: 0000-0001-6714-9043
              \and
              M.~Giroletti \at
              INAF Institute of Radio Astronomy, Bologna, Italy \\
              ORCID: 0000-0002-8657-8852
              \and
              C.~Goddi \at 
              Radboud University, Nijmegen, The Netherlands 
              \and
              J.-L. G\a'omez \at 
              Instituto de Astrof\a'isica de Andaluc\a'ia -- CSIC, Granada, Spain 
              \and
              J.~van~der~Gucht \at 
              Radboud University, Nijmegen, The Netherlands 
              \and
              J.C.~Guirado \at 
              Observatorio Astron\a'{o}mico, University of Valencia, Spain 
              \and
              Z.~Haiman \at 
              Department of Astronomy, Columbia University, New York, USA
              \and
         F.~Helmich \at 
         SRON Netherlands Institute for Space Research \\
         and \\
         Kapteyn Institute, University of Groningen, The Netherlands
         \and
         E.~Humphreys \at 
         ESO, Garching, Germany
         \and
         V.~Impellizzeri \at 
         Joint ALMA Observatory, Chile
         \and
         M.~Kramer \at 
         Max-Planck-Institut f\a"ur Radioastronomie, Bonn, Germany
         \and
         M.~Lindqvist \at 
         Department of Space, Earth and Environment, Chalmers University of Technology, Onsala Space
         Observatory, Sweden
         \and
         H.~Linz \at 
         Max Planck Institute for Astronomy, Heidelberg, Germany
         \and
         E.~Liuzzo \at
         INAF Institute of Radio Astronomy, Bologna, Italy
         \and
         A.~P.~Lobanov \at 
         Max-Planck-Institut f\a"ur Radioastronomie, Bonn, Germany
         \and
         Y.~Mizuno \at 
         Goethe-Universit\"at Frankfurt, Germany
         \and
         L.~Rezzolla \at 
         Goethe-Universit\"at Frankfurt, Germany
         \and
         F.~Roelofs \at 
         Radboud University, Nijmegen, The Netherlands
         \and
         E.~Ros \at 
         Max-Planck-Institut f\a"ur Radioastronomie, Bonn, Germany
         \and
         K.L.J.~Rygl \at 
         INAF Institute of Radio Astronomy, Bologna, Italy \\
         ORCID: 0000-0003-4146-9043
         \and
         T.~Savolainen \at
         Aalto University Department of Electronics and Nanoengineering, Finland \\
         and \\
         Aalto University Mets\a"ahovi Radio Observatory, Finland \\
         and \\
         Max-Planck-Institut f\a"ur Radioastronomie, Germany \\
         ORCID: 0000-0001-6214-1085
         \and
         K.~Schuster \at 
         IRAM, Grenoble, France 
         \and
         T.~Venturi \at 
         INAF Institute of Radio Astronomy, Bologna, Italy
         \and
         M.~C.~Wiedner \at 
         Observatoire de Paris, PSL University, Sorbonne Universit\'{e}, CNRS, LERMA, Paris, France 
         \and
         J. A.~Zensus \at
         Max-Planck-Institut f\a"ur Radioastronomie, Bonn, Germany
}

\date{Received: date / Accepted: date}

\maketitle

\begin{abstract}

This paper presents the ESA Voyage 2050 White Paper for a concept of TeraHertz Exploration and Zooming-in for Astrophysics (THEZA). It addresses the science case and some implementation issues of a space-borne radio interferometric system for ultra-sharp imaging of celestial radio sources at the level of angular resolution down to (sub-) microarcseconds. THEZA focuses at millimetre and sub-millimetre wavelengths (frequencies above $\sim$300~GHz), but allows for science operations at longer wavelengths too. The THEZA concept science rationale is focused on the physics of spacetime in the vicinity of supermassive black holes as the leading science driver. The main aim of the concept is to facilitate a major leap by providing researchers with orders of magnitude improvements in the resolution and dynamic range in direct imaging studies of the most exotic objects in the Universe, black holes. The concept will open up a sizeable range of hitherto unreachable parameters of observational astrophysics. It unifies two major lines of development of space-borne radio astronomy of the past decades: Space VLBI (Very Long Baseline Interferometry) and mm- and sub-mm astrophysical studies with ``single dish'' instruments. It also builds upon the recent success of the Earth-based Event Horizon Telescope (EHT) -- the first-ever direct image of a shadow of the super-massive black hole in the centre of the galaxy M87. As an amalgam of these three major areas of modern observational astrophysics, THEZA aims at facilitating a breakthrough in high-resolution high image quality studies in the millimetre and sub-millimetre domain of the electromagnetic spectrum.

\keywords{Radio interferometry \and VLBI \and mm- and sub-mm astronomy \and spaceborne astrophysics \and super-massive black hole}
\end{abstract}

\section{THEZA science rationale}

Astronomical advances are typically driven by technological advances and expansion of the parameter space made available for observing the Universe. Imaging has been and still is at the front line of astronomical research. There are two extreme ends one can consider. One extreme frontier is covered by large survey telescopes which charter large areas of the entire sky to make ever-more complete inventories of cosmic sources. Gaia has measured positions of billions of stars, revealing important aspects of Galactic structures. Euclid will map two billion galaxies to understand the structure of the Universe and the nature of dark energy. On the ground, telescopes like the Square Kilometre Array (SKA) promise to map billions of radio sources, including all powerful radio galaxies back to the beginning of the Universe. 

However, the other extreme end of the astronomical parameter space is higher resolution, high quality imaging. Rather than understanding the entire Universe at once, individual objects are studied with ever sharper vision. In optics, the Hubble Space Telescope has certainly changed our view of the Universe and made a big impact not only on science as such, but also on its perception by the general public.

Both approaches are necessary. After all, we cannot understand the large scales of the Universe, if we do not understand the objects that populate it and we cannot understand cosmic processes, if we only look at them with blurred vision.

Here we argue that the time is ripe for a major leap forward in astronomical resolution and image quality, providing us with orders of magnitude improvements in both areas and a stunning view of the most exotic objects in the Universe: black holes.

One of the major challenges of fundamental physics in the coming decades will be understanding the nature of spacetime and gravity, and black holes are at the centre of these challenges. Spacetime provides the underlying theatre within which the entire drama of our Universe unfolds. Gravity and the geometry of space are in principle well-described by the theory of General Relativity (GR). Yet this theory is still one of the biggest mysteries in theoretical physics. The presence of dark energy in the cosmos tells us that our understanding of spacetime is not complete and may require quantum corrections, which can even affect the largest scales. Similarly, the notion of Hawking radiation suggests that quantum theory and classical black holes seem to be incompatible. However, after many decades of research the unification of GR and quantum theory is still a major problem and perhaps experiments may now need to lead the way.

Fortunately, we are now entering an era when experimental tests of gravity -- even under the most extreme conditions -- are becoming possible. The nature of dark energy is targeted by large scale surveys, as mentioned before. The Square Kilometre Array (SKA) will have the ability to detect and measure new pulsar systems to significantly improve on existing tests of GR. New X-ray missions like Athena will allow us to do spectroscopic measurements of hot gas orbiting black holes and gravitational wave experiments like LISA or the Einstein Telescope will provide detailed measurements of the dynamical nature of spacetime. Hence, one could claim that the past century was the century of particle physics and astrophysics, while this century promises to be the century of experimental spacetime physics - the ultimate synthesis of the two. 

\begin{figure}[t]
  \centering
\includegraphics[width=0.99\textwidth,angle=0]{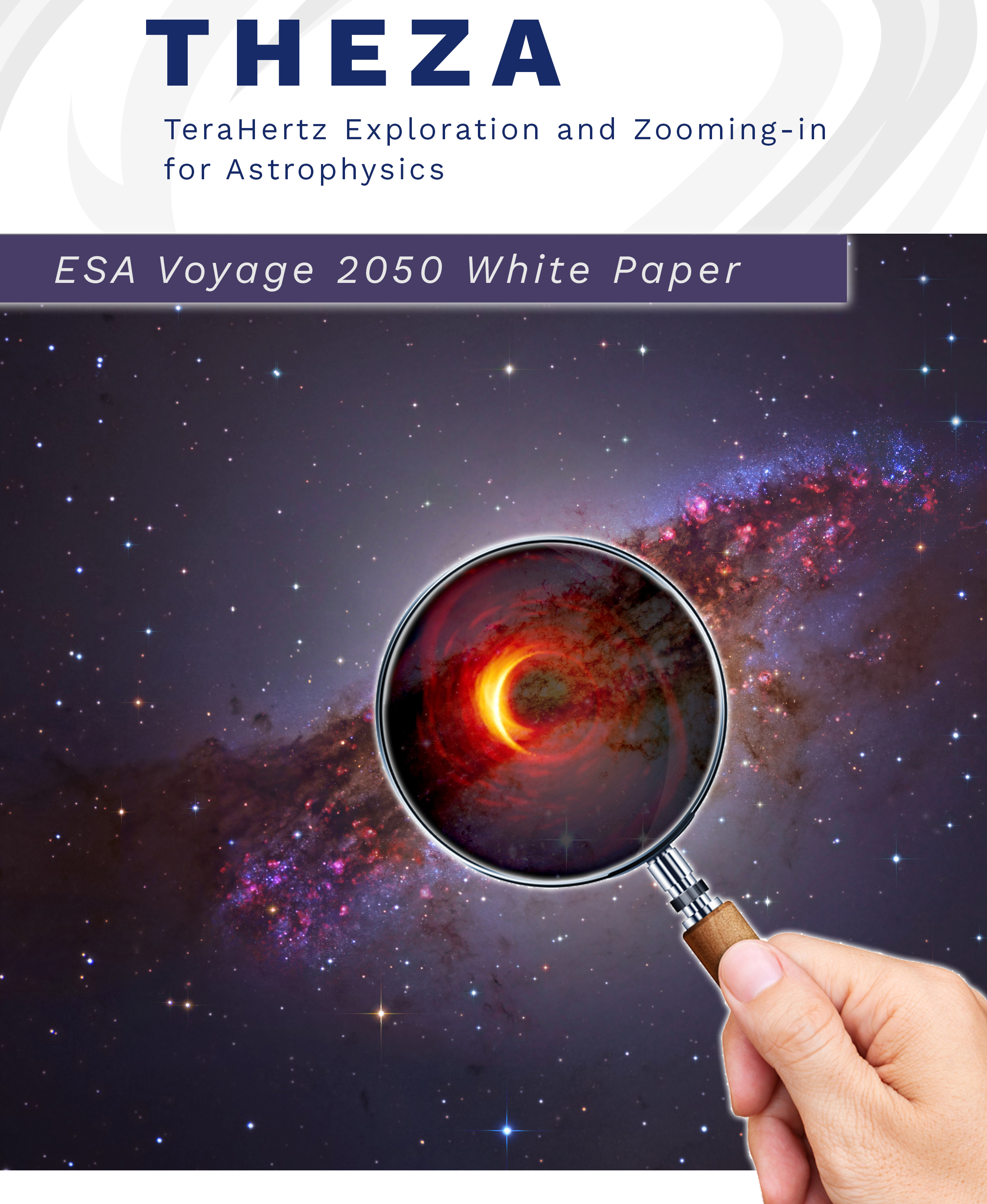}
\caption{Zooming into the central area of a galaxy containing a SMBH. Artwork by Beabudai Design, the background -- Cen~A free stock image, a simulated image of a central area of a galaxy containing a supermassive black hole \citep{Moscibrodzka2014}.
}
   \label{fig:THEZA-cover}   
\end{figure}

However, of all the techniques, there is only one that allows one to make actual images of black holes and the astrophysical processes surrounding them: interferometry.  Optical interferometry with ESO's Gravity experiment has imaged the motion of gas around the Super-massive Black Hole (SMBH) in the centre of Milky Way \citep{Gravity-CollaborationAbuterAmorim2018a,Gravity-CollaborationAbuterAmorim2018b} and Very Long Baseline Interferometry (VLBI) in the radio domain has imaged the innermost structures of jets and recently even the black hole shadow in the centre of the radio galaxy M87 \citep{EHT2019I,EHT2019II,EHT2019III,EHT2019IV,EHT2019V}. These new results are quantum leaps and they only mark the beginning. By going into space, the imaging resolution and fidelity can be enhanced enormously, and we will be able to see black holes and their immediate environment in unprecedented detail and quality (Fig.~\ref{fig:THEZA-cover}). Any physical and astrophysical theory of black holes and their activity will need to be able to predict in detail how they look.

In this paper we discuss the promises of space-based VLBI for the study of black holes within the concept dubbed TeraHertz Exploration and Zooming-in for Astrophysics (THEZA). Of course, once one has access to the extreme resolution provided by Space VLBI, other than SMBH imaging science cases will also become possible with the same technology, which we also briefly touch upon.

\section{Science and technology heritage}

\subsection {Space VLBI}\label{Subsec_SVLBI}

For more than half a century, since its first demonstrations in 1967, Very Long Baseline Interferometry (VLBI) holds the record in sharpness of studying astronomical phenomena. This record reaches angular resolutions of milliarcseconds and sub-milliarcseconds at centimetre and millimetre wavelengths owing to the ultimately long baselines, comparable to the Earth's diameter. However, the hard limit of VLBI angular resolution defined by the size of the Earth does not allow us to address astrophysical phenomena requiring even sharper sight. Not surprisingly, soon after the demonstration of first VLBI fringes with global Earth-based systems, a push for baselines longer than the Earth's diameter materialised in a number of design studies of Space VLBI (SVLBI) systems.

Over the past decades, several dozens of various SVLBI concepts have been presented with widely varying depth of development and level of detail \citep{Gurvits2018,Gurvits2019}. They paved the way for the first SVLBI demonstration in the middle of the 1980s by the TDRSS Orbital VLBI experiment \citep{Levy+1986}, and two dedicated SVLBI missions, VSOP/HALCA launched in 1997 \citep{Hirabayashi+1998} and RadioAstron launched in 2011 \citep{Kardashev+2013}.

\begin{figure}[h]
  \centering
\includegraphics[width=0.47\textwidth,angle=90]{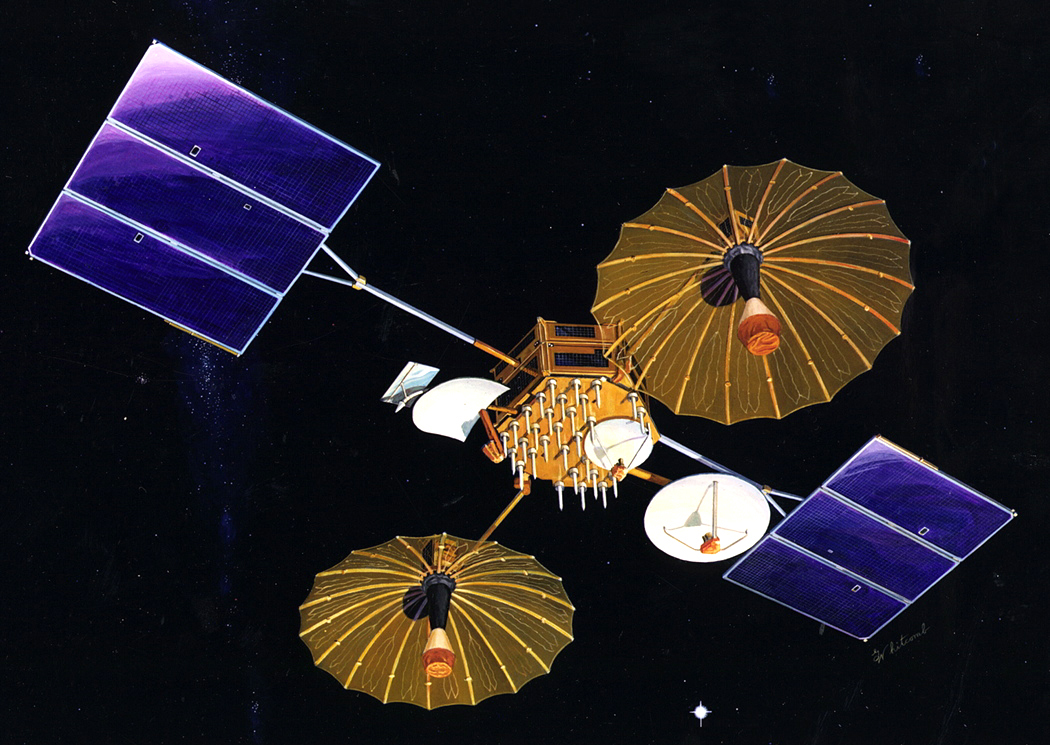}
\includegraphics[width=0.47\textwidth,angle=90]{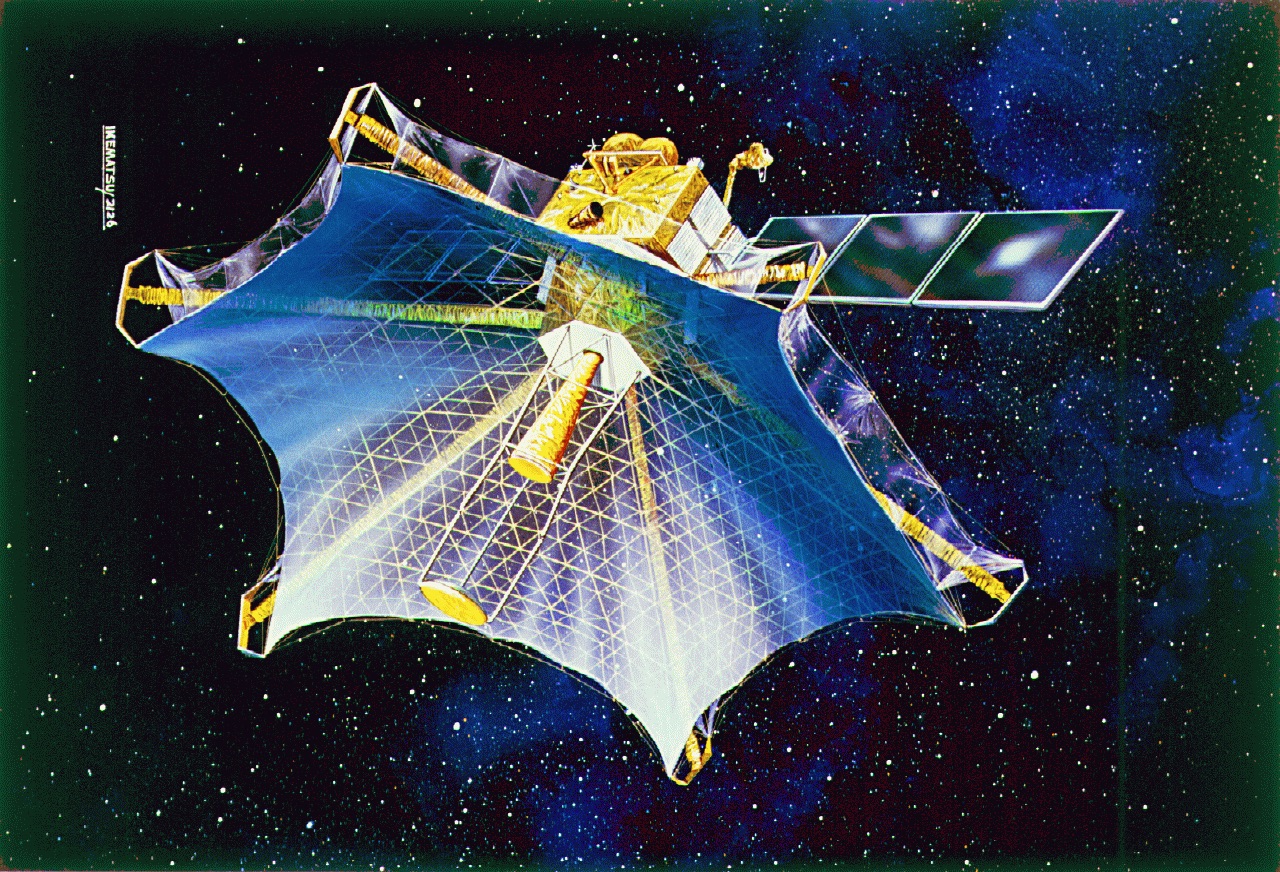}
\includegraphics[width=0.47\textwidth,angle=90]{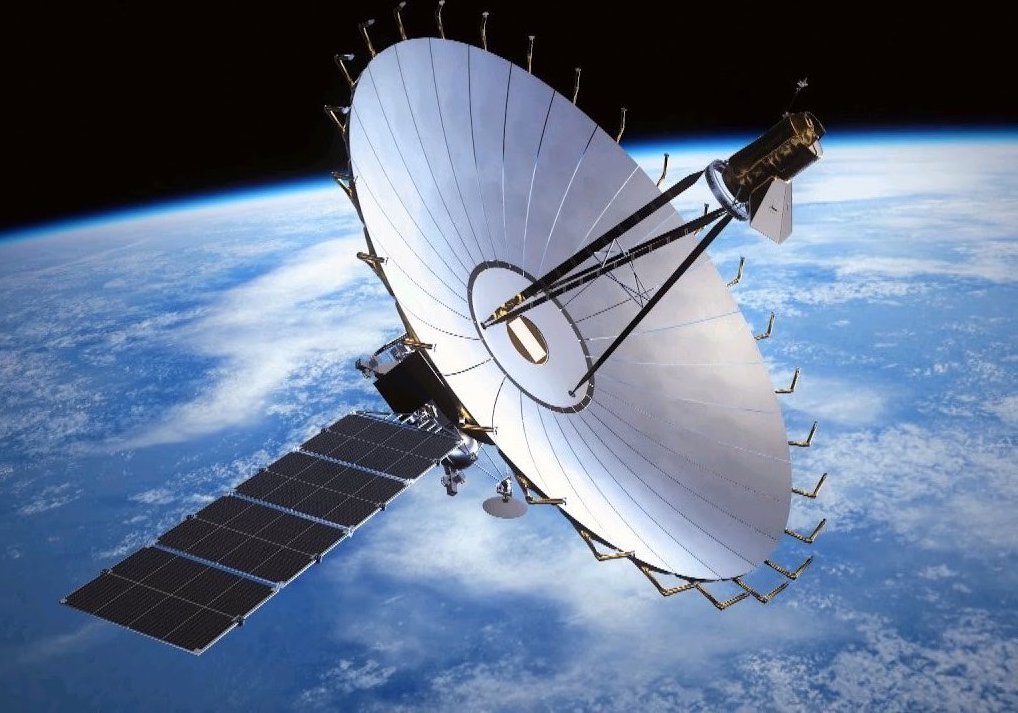}
\caption{Artist's impressions of (\textit{left to right}): Tracking and Data Relay Satellite System (TDRSS) spacecraft, picture credit: NASA; Highly Advanced Laboratory for Communication and Astronomy (HALCA) of the VLBI Space Observatory Program (VSOP), picture credit: JAXA; Spektr-R spacecraft of the RadioAstron mission, picture credit: Lavochkin Scientific and Production Association.}
   \label{fig:SVLBI}   
\end{figure}

First VLBI ``fringes'' on baselines longer than the Earth's diameter were obtained with NASA’s geostationary spacecraft of the Tracking and Data Relay Satellite System (TDRSS, shown in the left panel of Fig.~\ref{fig:SVLBI}) in 1986 \citep{Levy+1986}. This was a very efficient example of ad hoc use of existing orbiting hardware not designed originally for conducting SVLBI observations. The main outcome of several TDRSS observing campaigns was two-fold. First, the very concept of getting coherent interferometric response (the so called interferometric ``fringes'') on baselines longer than the Earth's diameter was proven experimentally. Second, observations of two dozen of the strongest Active Galactic Nuclei (AGN) at 2.3 GHz in 1986--87 \citep{Levy+1989,Linfield+1989}, and in a dual-frequency mode at 2.3~GHz and 15~GHz in 1988 \citep{Linfield+1990} provided indications that at least some extragalactic sources of continuum radio emission were more compact and therefore brighter than expected. These milestones supported growing momentum for the first generation of dedicated Space VLBI missions. 

The VSOP/HALCA (Fig.~\ref{fig:SVLBI}, middle) operated in orbit in the period 1997--2003. Its science heritage is summarised in \citet{Hirabayashi+2000a,Hirabayashi+2000b} and \citet[Parts 3-4]{Hagiwara+2009}. The RadioAstron mission (Fig.~\ref{fig:SVLBI}) was operational in the period 2011--2019. Its science outcome is still to be worked out; some preliminary summaries are presented by \citet{NSK+YYK17} and in the Special Issue of Advances in Space Research \citep{JASR2019}. A very brief list of major legacy achievements of VSOP and RadioAstron missions with associated references is given in \citet{Gurvits2018}. The major qualitative result of the first generation SVLBI missions, VSOP and RadioAstron can be expressed as follows: sub-milliarcsecond angular scale in continuum (AGN, pulsars) and spectral line (maser lines of hydroxyl and water molecules) sources is consistent with the current general understanding of the astrophysics of these sources, but various enigmatic details require further in-depth studies with similar or sharper resolution and higher sensitivity. 

The first generation SVLBI era has come to completion in 2019. It has provided a solid proof of concept of radio interferometers exceeding the size of Earth and serves as a stepping stone toward future advanced SVLBI systems presented in this White Paper.

\subsection {Millimetre and sub-millimetre space-borne radio astronomy}

Over the last three decades, developments for ground and space-borne millimetre and sub-millimetre heterodyne detection instruments have generated important synergies. The SWAS \citep{Melnick+2000}, Odin \citep{Frisk+2003} and NASA Earth Observing System Microwave Limb Sounder \citep{Waters+2006} orbital missions were all equipped with room temperature or cooled Schottky receivers. One of the most important steps in advancing modern heterodyne instruments was the development of cryogenic systems with superconducting mixer elements (SIS). In particular, such technology was essential not only for  achieving quantum limited noise but could also work with much lower local oscillator (LO) power, a condition permitting very compact and versatile solid state LO generators. The complex Herschel HIFI instrument adopted this technology for the on-board instrumentation enhanced by stringent quality assurance schemes \citep{deGraauw+2010}. The success of HIFI showed that SIS and hot-electron bolometer (HEB) technology is very well suited for space application. 

This technology path found its way into advanced mm/sub-mm systems both for Earth-based facilities (e.g., the ALMA and IRAM's NOEMA interferometers). The recent study for Origins (one of the four NASA flagship candidates for the next Decadal) included a thorough look at the heterodyne needs for the HERO (HEterodyne Receiver for Origins) onboard instrument \citep{Wiedner+2018}. This study showed that achieving a high Technology Readiness Level (TRL) sufficient for implementation within the timeframe of the 2020 US Decadal survey is possible and no showstoppers are foreseen. This also holds for direct detectors, as studied for SPICA and Origins, and in a different incarnation for the ESA's Cosmic Vision L2 mission Athena (the XIFU instrument). The Transition Edge Sensors (TES) and Kinetic Induction Detectors (KID) devices provide ultimate sensitivity and, if needed, high multiplexing capabilities. The technology required for these detectors has been developed, in particular, at the SRON, The Netherlands. Finally, very powerful broadband digital signal processing is now available which enables data processing with extreme efficiency, while power consumption for these digital operations is expected to decrease within the coming years to sufficiently low levels for space application.

\begin{figure}[h]
  \centering
\includegraphics[width=0.13613\textwidth,angle=0]{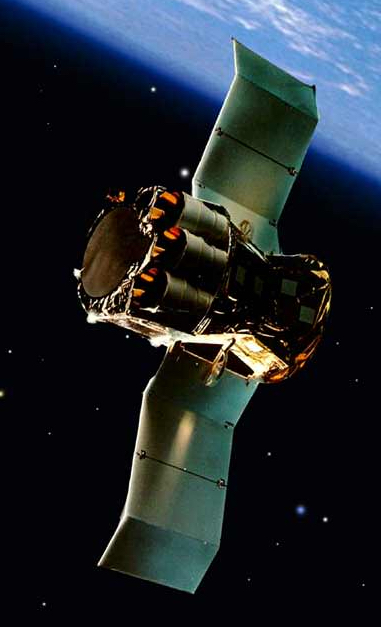}
\includegraphics[width=0.14850\textwidth,angle=0]{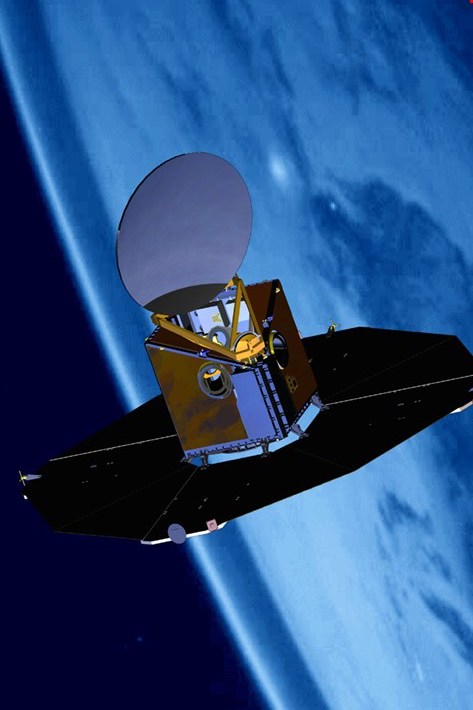}
\includegraphics[width=0.1705\textwidth,angle=0]{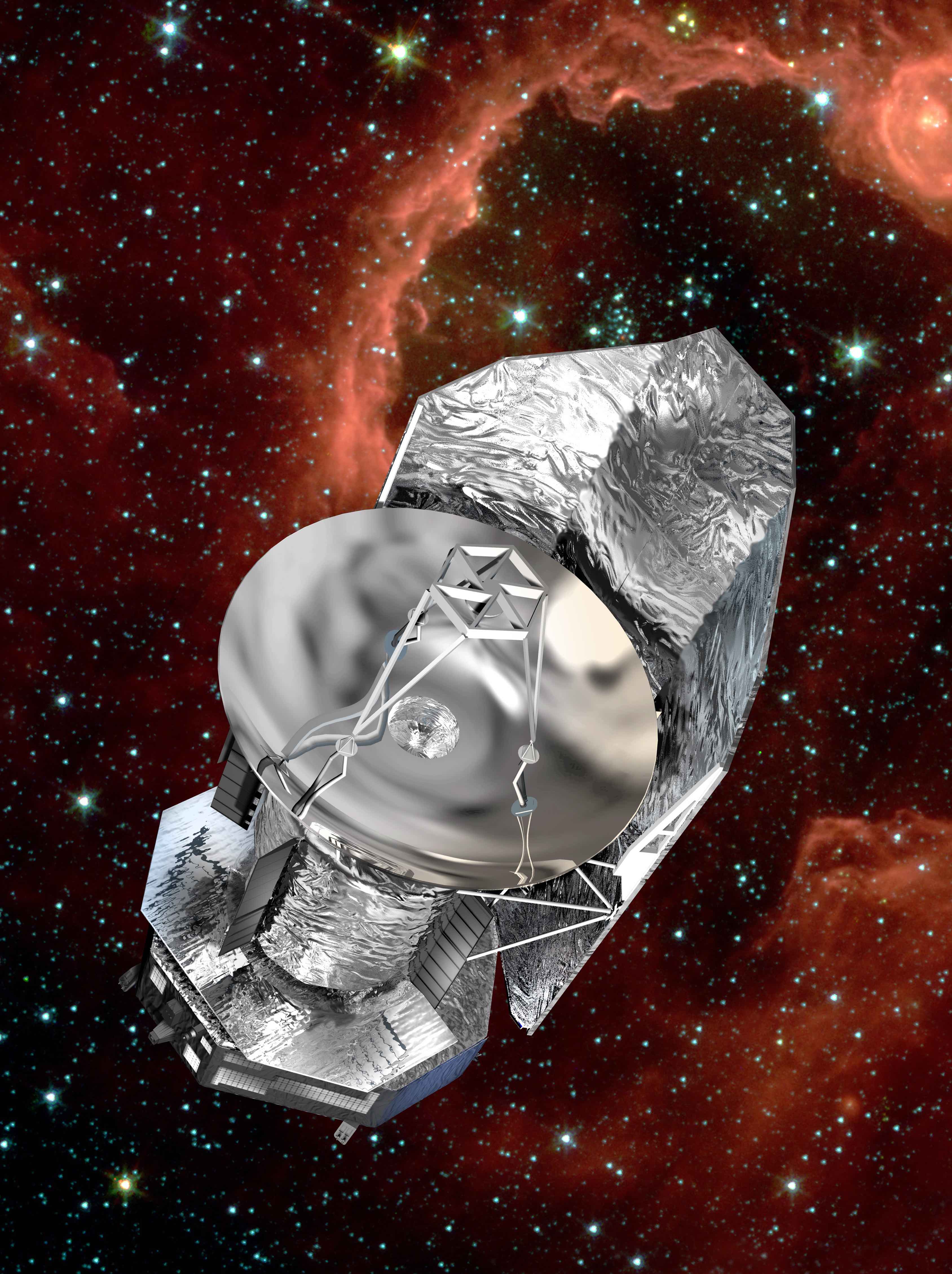}
\includegraphics[width=0.16418\textwidth,angle=0]{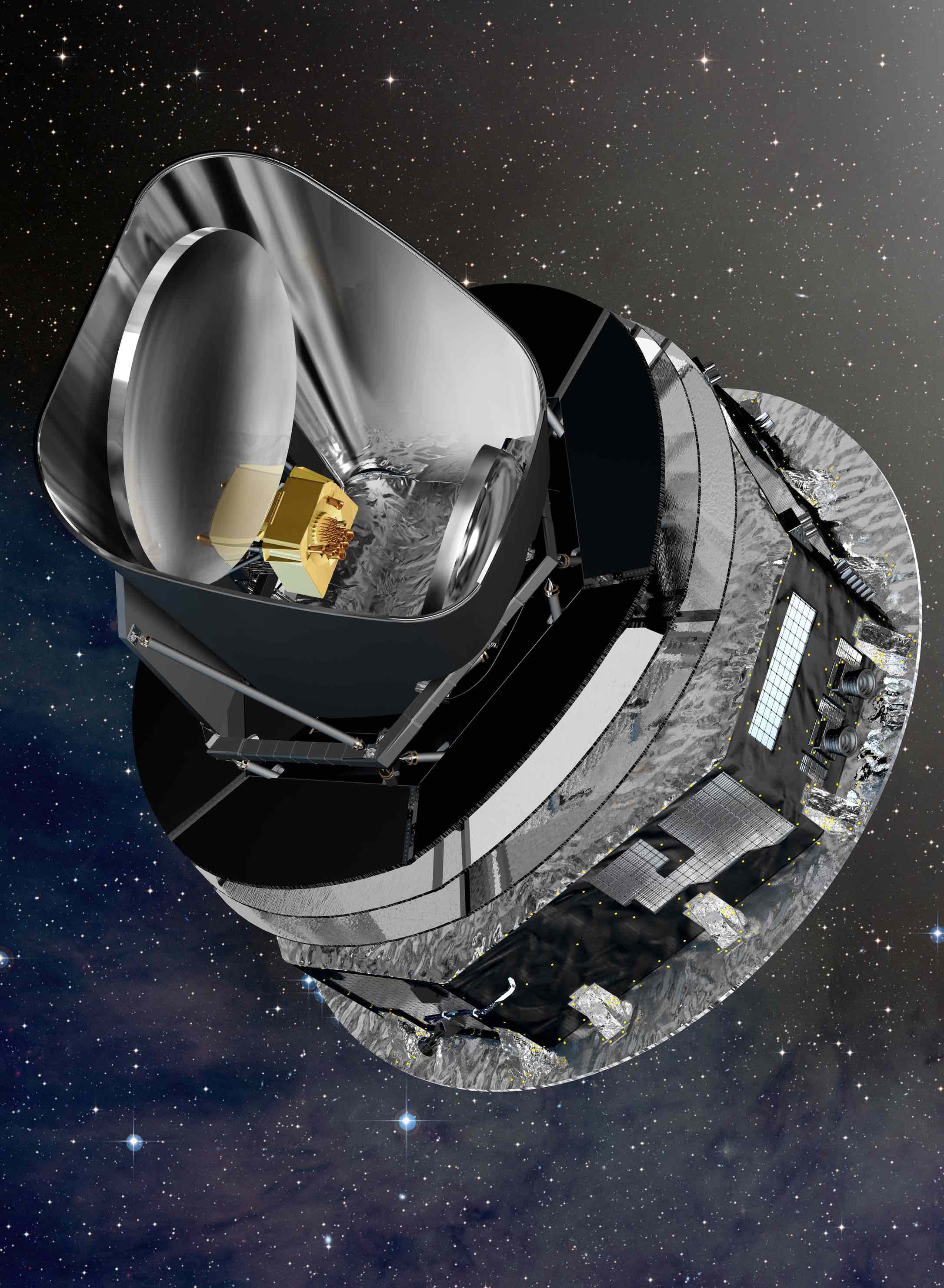}
\includegraphics[width=0.1705\textwidth,angle=0]{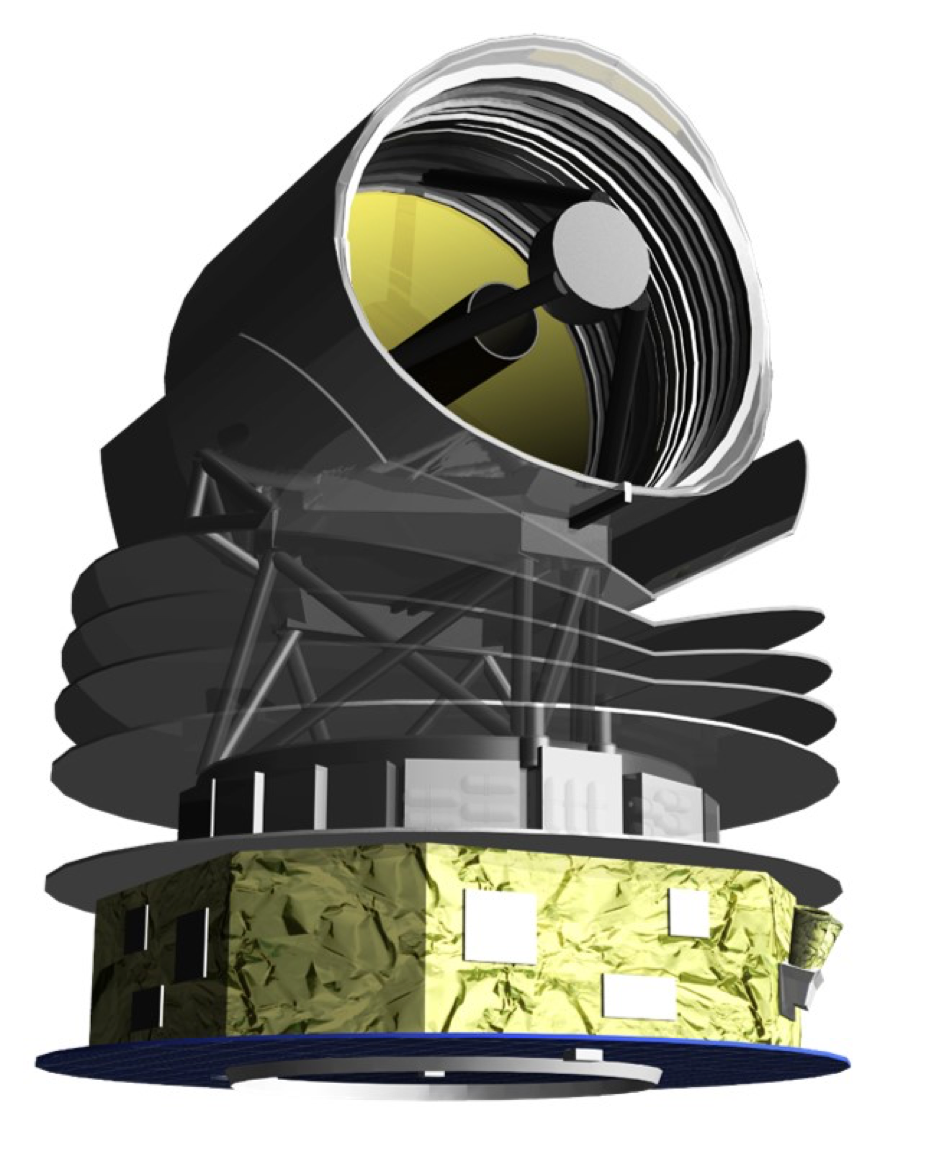}
\includegraphics[width=0.1705\textwidth,angle=0]{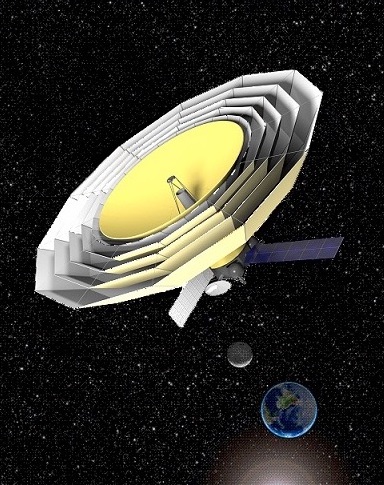}
\caption[ ]{Artist's impressions of (\textit{left to right}): \protect\\
      \begin{tabular}{cl}
   -- & Sub-Millimetre Wave Astronomy Satellite (SWAS), picture credit: NASA, USA; \\
   -- & Odin satellite, picture credit: the Swedish National Space Agency, Sweden; \\
   -- & Herschel spacecraft, picture credit: ESA; \\
   -- & Planck spacecraft, picture credit: ESA; \\
   -- & a concept of the SPace IR telescope for Cosmology and Astrophysics (SPICA), \\
      & picture credit: SRON, The Netherlands; \\
   -- & a concept of the Millimetron mission, picture credit: Astro Space Center, Russia.  
      \end{tabular}}
   \label{fig:mm-missions}   
\end{figure}

Other important improvements in cryogenic heterodyne instruments have taken place and are still moving forward. Instantaneous bandwidth is now an orders of magnitude larger than 20 years ago also due to the greatly improved performance of cryogenic IF amplifiers. 

The Herschel instruments were passively cooled to 80~K and the instruments were kept cold through boiling of Helium gas in a large cryostat. With the developments of cryocoolers for Planck, the cryocoolers of Hitomi and the current developments for Athena and SPICA, we expect that any mission needing low temperatures can rely on closed cycle cryocooling rather than liquid Helium cryostats. Planck also showed the value of the V-groove system, currently employed in Ariel and studied for the SPICA satellite. All in all, the available heritage of mm/sub-mm space missions and their instrumentation and ongoing developments assure beyond doubt that technical specifications for THEZA can be met within the Voyage 2050 time-frame. Fig.~\ref{fig:mm-missions} presents artist's impressions of four completed missions (SWAS, Odin, Herschel and Planck) as well as two concepts (SPICA and Millimetron) which can be seen as stepping stones toward a mission addressing the THEZA concept. 

\subsection{Ground-based radio astronomy arrays: EVN, Global VLBI, GMVA, EHT}

With a collecting area comparable to that of the first phase of the mid-frequency Square Kilometre Array (SKA1-MID), today the European VLBI Network (EVN\footnote{https://www.evlbi.org/home, EVN site accessed 2020.12.24.}) is a joint facility of independent European, African, Asian and North American radio astronomy institutes with thirty-two radio telescopes throughout the world and operating in a range of wavelengths going from 92\,cm to 0.7\,cm. The EVN offers (sub-)milliarcsecond angular resolution with $\mu$Jy sensitivity in those bands with best performances (21/18\,cm and 6/5\,cm). It is the most sensitive regular VLBI array adopting open skies policies\footnote{Open skies policy provides open access to astronomical facilities for the world-wide scientific community free of charge on the basis of peer-reviewed observing proposals.}. Since the start of its operations in the early 1980s, the EVN has been undergoing continuous development to match the scientific requirements, the most relevant being: the increased recording rate at each telescope from the initial 4\,Mbps to 2\,Gbps at present; the increased number of member observatories, which has brought the number of antennas from 5 to the current 32, with an amazing jump in the image fidelity; and the real-time operations with a subset of the array through the e-VLBI, whereby the recorded signal is transferred to the EVN correlator at JIVE (the Joint Institute for VLBI European Research Infrastructure Consortium) through fibre link connection, a feature which has made the e-EVN array an SKA pathfinder.

Owing to the broad frequency range and high sensitivity offered to the community, the EVN delivers outstanding science in almost any astrophysical area, from the distribution of dark matter through gravitational lensing, to the origin of relativistic jets in AGN and their evolution with cosmic time, to stellar evolution -- from the pre-main sequence stage to the post-asymptotic giant branch stage -- and to the successful detection of exoplanets. More recently, it has become clear that VLBI can uniquely contribute to what is now referred to as the {\it science of transient phenomena}, a broad range of very energetic phenomena covering fast radio bursts, gamma-ray bursts, tidal disruption events, and follow-up of electromagnetic counterparts to gravitational wave events, thanks to the combination of superb angular resolution and sensitivity, and microarcsecond precision localisation.

While the EVN as such has no frequency overlap with the THEZA concept presented here, it is the vast area of complementary science that makes the EVN a highly synergistic to the THEZA science. Moreover, the most important synergy between EVN and THEZA is in human capital: most members of the THEZA Team are active EVN functionaries and users. In short, the EVN is a major ``breeding ground'' for ideas, technologies and know-how behind the THEZA initiative. 

The already superb angular resolution of the EVN can be sharpened by observing at shorter wavelengths, as has been achieved by the Global Millimetre VLBI Array (GMVA\footnote{https://www3.mpifr-bonn.mpg.de/div/vlbi/globalmm/, GMVA site accessed 2020.12.24.}), another world-spanning array consisting of sixteen antennas operating at 3.5\,mm (85--95 GHz band). The array may be further expanded with the addition of the phased ALMA and the Korean VLBI Network. Further potential enlargement of the GMVA will involve the Greenland Telescope and the Mexican Large Millimetre Telescope.  Exploration of the inner jet regions of nearby AGN and blazars is one of the main scientific drivers for 3\,mm GMVA observations. 

The experience acquired with the GMVA, in operation since the early 2000s, for which observations, calibration, and data analysis pose several challenges due to the weather dependence and the heterogeneity of antennas in the array, has paved the way to the  Event Horizon Telescope (EHT\footnote{https://eventhorizontelescope.org, EHT site accessed 2020.12.05.}), which has pushed the angular resolution and observing frequencies to the limits of the current capabilities of ground-based VLBI. Imaging the black hole shadow of the nearby galaxy M\,87 has been one of the most remarkable achievements of the VLBI technique to date (\citet{EHT2019I} and references therein), and this has been possible owing to 50 years of a worldwide collaborative effort based on scientific, technological, and human capital investment.

The EHT is the ultimate development of the Earth-based VLBI in terms of the wavelength and geometry of the array. It serves as a benchmark for the THEZA concept presented here. The next steps forward in development of VLBI at frequencies above 100~GHz and into the THz regime are necessitated by the astrophysical motivations. They promise transformational science results and require VLBI systems above the Earth's atmosphere. This is the essence of the THEZA concept. 

\section{THEZA science}
\label{Section_Science}

\subsection{Event horizon in SMBH: physics and cosmology}

In 2019, after decades of developmental efforts from a global collaboration of scientists, the EHT collaboration presented the first image of a black hole \citep{EHT2019I}. The EHT image, revealing the supermassive black hole (SMBH) in M87, was captured using a global VLBI network operating at a wavelength of 1.3 mm. It is formed by light emitted near the black hole and lensed towards the photon orbit of the 6.5 billion solar mass black hole at the galaxy's core. The combination of light bending in curved spacetime and absorption by the event horizon leads to a characteristic black hole shadow embedded in a light ring that was predicted by \citet{FalckeMeliaAgol2000} to be observable with mm-wave VLBI. The results constrain the object to be more compact than the photon orbit, at a size $\le 3/2 R_{\rm S}$, where $R_{\rm S}$ is the Schwarzschild radius. This is comparable to the compactness currently probed by LIGO/VIRGO with gravitational waves, but on mass scales eight orders of magnitude larger.

The capability to image black holes on event horizon scales enables entirely new tests of General Relativity (GR) near a black hole,  e.g., \citet{JohannsenPsaltis2010a}, and opens a direct window into the astrophysical processes that drive accretion onto a black hole and formation of relativistic jets \citep{MoscibrodzkaFalckeShiokawa2016a}. Jets are a major source of energy output and high-energy emission for black holes across all mass scales. Their origin and formation is of fundamental importance to high-energy astrophysics and happens on event horizon scales.

For precise tests of GR and time-domain studies of accretion flows and jet formation, we therefore need sharper angular resolution, higher observing frequencies, and faster and more complete sampling of interferometric baselines.

The angular resolution of ground-based VLBI is approaching fundamental limits. Interferometer baseline lengths are currently limited to the diameter of the Earth, imposing a corresponding resolution limit for ground arrays of $\sim$22\,$ \mu \mathrm{as}$ at an observing frequency of 230 GHz. Observations at higher frequencies can improve the angular resolution but become increasingly challenging because of strong atmospheric absorption and rapid phase variations, severely limiting the number of suitable ground sites and the windows of simultaneous good weather at many global locations. The  fixed telescope locations also limit the number of Fourier-components of the image that can be sampled and hence the image quality.  Going into space would overcome these limitations and open new scientific possibilities for horizon-scale studies in SMBH.

Measuring the shape and size of the shadow and surrounding lensed photon ring in M87* and Sgr~A* provides a null hypothesis test of GR \citep{PsaltisOzelChan2015a}. Better images allow one to measure spin, test the no-hair theorem, measure the structure of the spacetime, and test for the possibilities for black hole alternatives, e.g., \citet{MizunoYounsiFromm2018a}.   

The big advantage of SMBH imaging, with respect to gravitational wave experiments, is that the sources are stable and their parameters can be determined independently with ever better accuracy. Extraction of parameters is achieved by comparing matched image templates from GRMHD simulations \citep{EHT2019V,EHT2019VI}.  For example, in Sgr~A*, the mass and distance are already well-measured within $\sim 1\%$ \citep{Gravity-CollaborationAbuterAmorim2018a}. Therefore the precision of testing GR for Sgr~A* is already limited by the fidelity of observing data and the ability to extract the emission corresponding to the black hole photon orbit and interior shadow. With the current resolution the  sharply delineated lensed photon ring and more extended lensed emission structures are still blurred together, making detailed tests or spin measurements almost impossible.  

\begin{figure}[t]
    \centering
    \includegraphics[width=\textwidth]{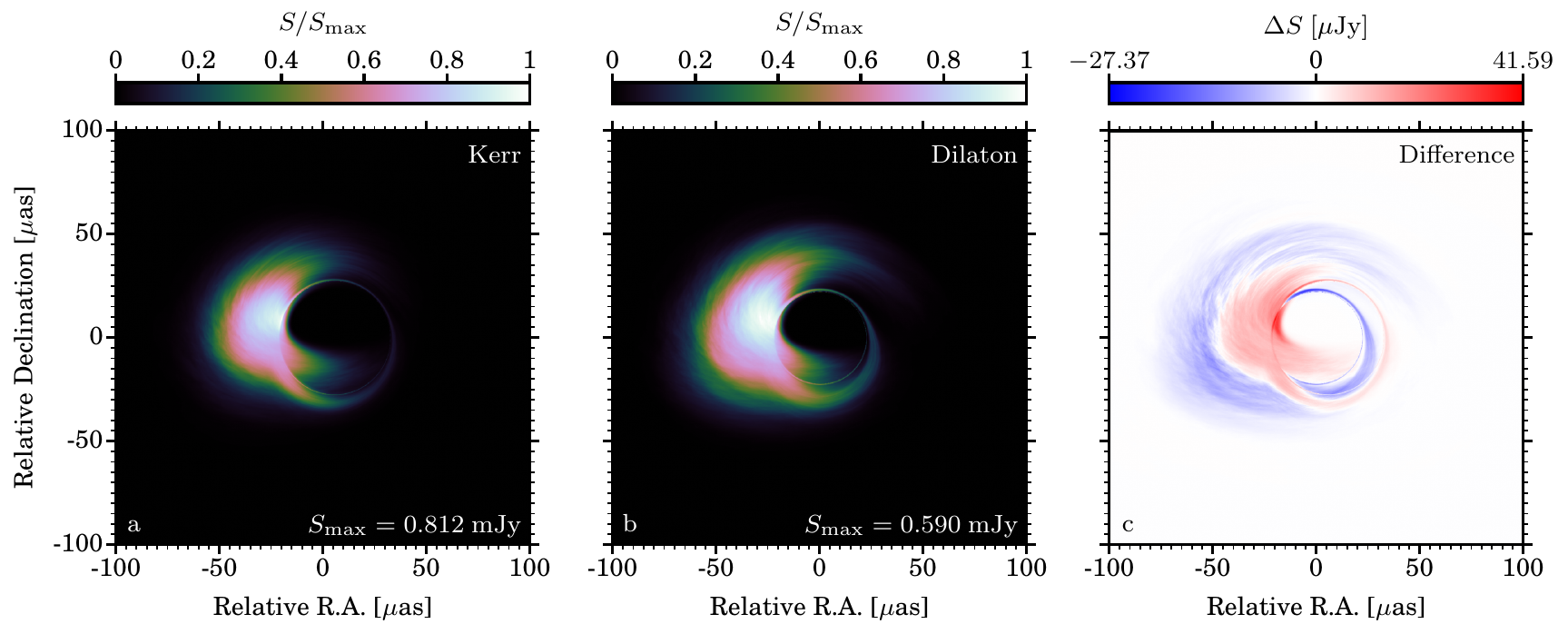}
    \caption{\textit{Left:} simulated image at 230 GHz from a GRMHD simulation of accretion onto a Kerr black hole. \textit{Middle:} the same for a non-rotating dilaton black hole. \textit{Right:} the difference between the two images. Figure from \citet{MizunoYounsiFromm2018a}.
}
    \label{fig:dilaton}
\end{figure}

Higher angular resolution by Space VLBI will allow more precise measurement of the shadow size and shape and increased dynamic range will improve image fidelity. Space VLBI observation allows us to extract the thin, lensed photon ring feature from the more diffuse surrounding emission. Such high-resolution images of the black hole shadow will allow us to constrain the physics of the black hole itself. Fig.~\ref{fig:dilaton} shows that the difference between accretion onto a Kerr black hole and a dilation black hole, which represents a modification of general relativity, becomes apparent in the size and shape of the photon ring at a resolution of 5-10 $\mu$as. The resolution of the ground-based EHT is not sufficient to distinguish between these cases based on reconstructed images \citep{MizunoYounsiFromm2018a}, but Space VLBI concepts will be able to reach the required resolution (see also Section \ref{sec:missionprofiles}).

\begin{figure}[ht]
    \centering
    \includegraphics[width=0.35\textwidth]{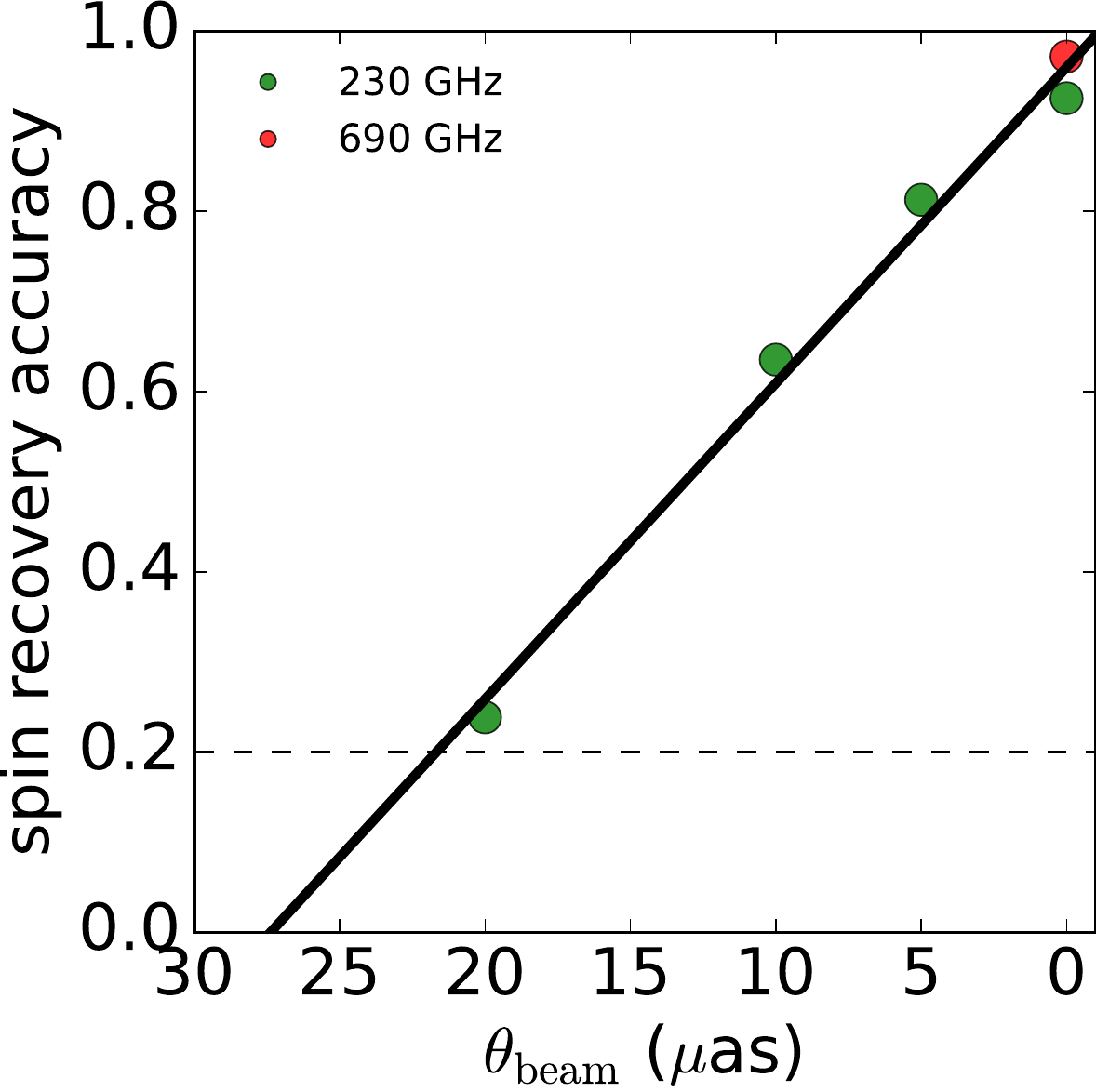}
    \includegraphics[width=0.53\textwidth]{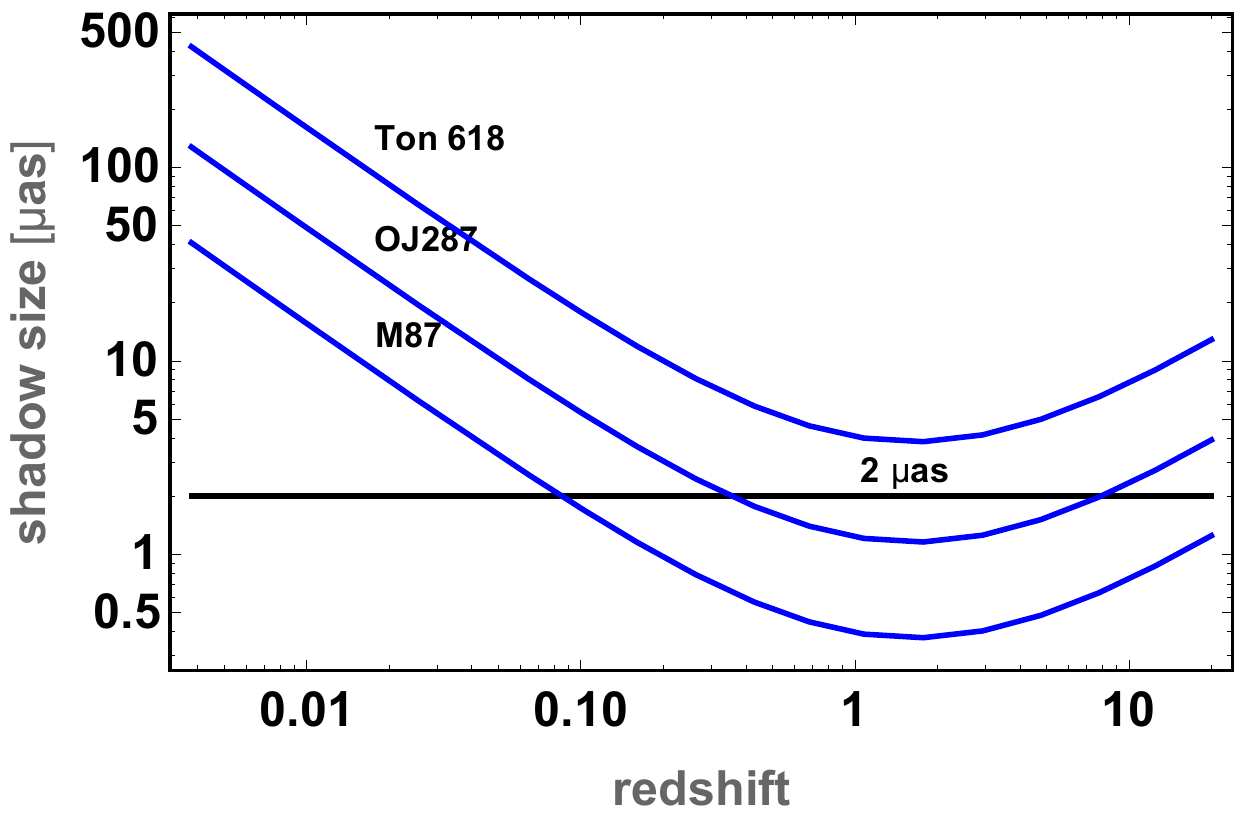}
    \caption{\textit{Left:} Black hole spin recovery accuracy from GRMHD simulations by a machine learning classifier network as a function of telescope beam size, where $1$ means perfect accuracy and $0.2$ means random guesses as five black hole spin values between $-1$ (maximal retrograde spin) and 1 (maximal prograde spin) were considered in this analysis. At the EHT resolution at 230~GHz ($\sim$20\,$\mu$as), the spin recovery accuracy is low. Only at high resolutions of around 5~$\mu$as, as could be attainable with Space VLBI, is the classifier network capable of recovering the correct spin value. \textit{Right:} Size of black hole shadows for three different mass ranges of SMBHs as function of redshift. We use as examples: M87 ($6.4\times10^9 M_\odot$), OJ287 ($2\times10^{10} M_\odot$), and TON618 ($6.6\times10^{10} M_\odot$), but allow them to be at any distance.}
    \label{fig:spin}
\end{figure}

Another quantity that becomes measurable at this resolution is black hole spin. For a Kerr black hole, the effect of black hole spin on the shadow size is limited to about 4\% \citep{JohannsenPsaltis2010b}, which is impossible to measure with ground-based VLBI. Fig.~\ref{fig:spin} shows that using machine learning techniques, the black hole spin starts  becoming measurable at 5 $\mu$as rsolution at 230~GHz \citep{2020A&A...636A..94V}. High-frequency imaging at 690 GHz or higher will, therefore, allow for stronger constraints on the spin.

%{\bf HF: This is a key argument for why we need more resolution - please explain a little better. Figure is not intuitive.}

The relatively small mass of Sgr~A*, $4.1 \times 10^6 \, \mathrm{M}_\odot$ \citep{Gravity-CollaborationAbuterAmorim2018a}, results in correspondingly short dynamical timescales (of order ten minutes) for the system. The current EHT array lacks sufficient baseline coverage to form images on these timescales. The rapid baseline sampling of an orbiter is necessary to recover the complex structure in an evolving accretion flow. Movies of Sgr~A* by multiple snapshot images will clarify the nature of coherent orbiting features such as "hotspots" \citep{Gravity-CollaborationAbuterAmorim2018b} and the origin of the flaring events observed in many wavebands, e.g., \citet{MarroneBaganoffMorris2008}. 

Magnetic fields play an important role in accretion and jet formation. The magneto-rotational instability (MRI,  \citet{BalbusHawley1998}) in accretion disc is thought to transport angular momentum and drive accretion onto the central black hole. Magnetic fields can also cause instabilities and flaring on horizon scales \citep{TchekhovskoyNarayanMcKinney2011}. Polarimetric imaging of polarised synchrotron radiation observed by Space VLBI can reveal the structure and dynamics of magnetic fields near the horizon. It will allow us to probe the magnetic field degree of ordering, orientation, and strength through Faraday rotation studies. Power spectral analysis will provide information on the turbulent accretion flow on very fine spatial scales, for the first time observationally  testing our understanding of MRI and angular momentum transport in the inner part of accretion disc \citep{BalbusHawley1998}.

Finer angular resolution also provides access to additional targets with spatially resolved black hole shadows. 
At $\sim 5 \, \mu \mathrm{as}$ angular resolution, the number of known nearby SMBHs that are expected to resolve the black hole shadow will increase also from two (Sgr~A* and M87*) to six (with the addition of M85, Cen~A, M104, IC1459), allowing more robust tests. At $\sim 2 \, \mu \mathrm{as}$ also IC4296 and M81 become accessible, but more importantly also large SMBHs at cosmological distances, e.g. something like OJ287 at $z=0.3$ or SMBH monsters such as TON618 at all redshifts.

So, an order of magnitude increase in resolution will provide at least an order of magnitude more sources for black hole shadow tests and access to the jet launching regions of hundreds to thousands of powerful AGN. 

\begin{figure}[ht]
    \centering
    \includegraphics[width=0.75\textwidth]{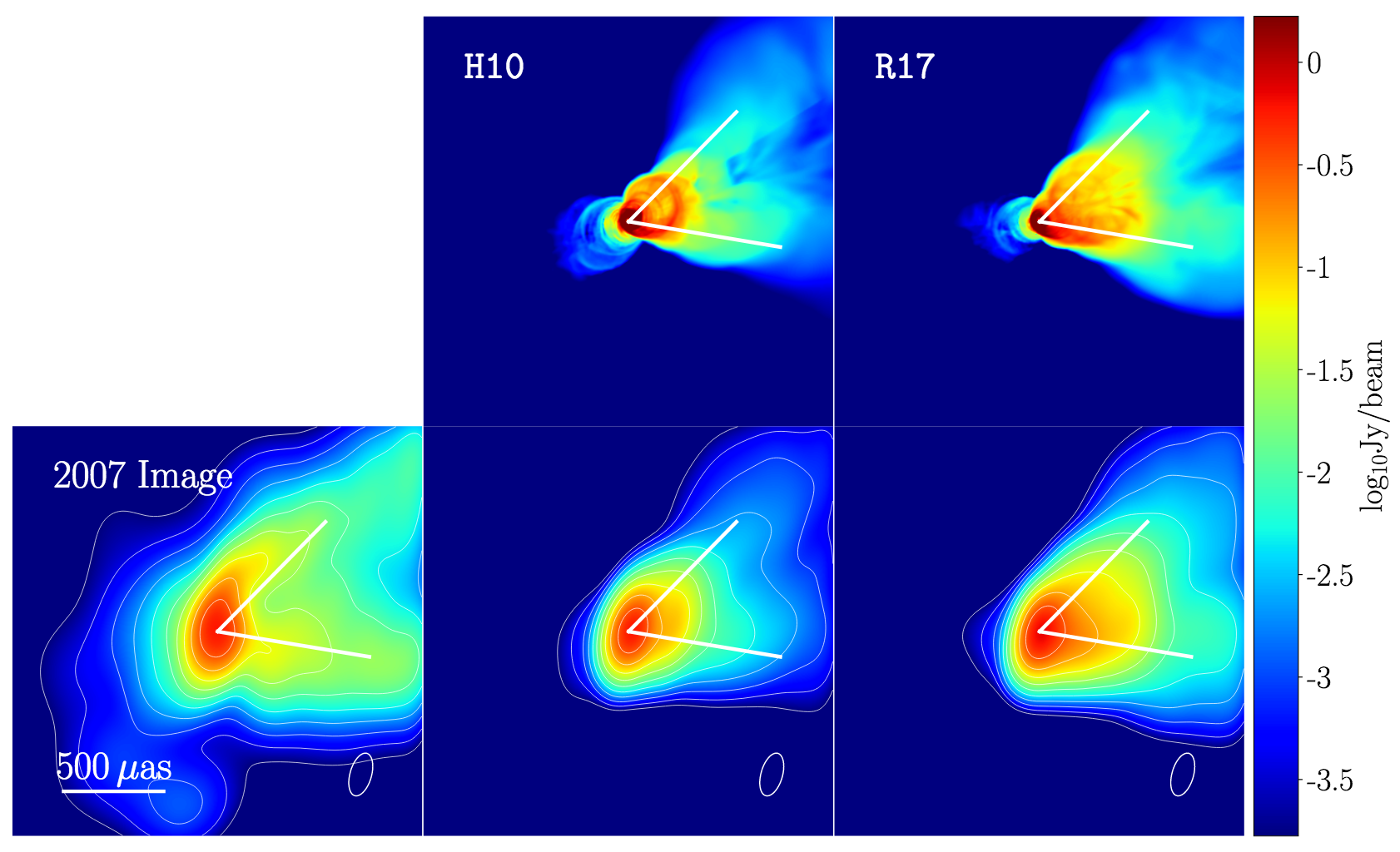}
    \caption{VLBA image of M87 at 43~GHz (left), compared to  simulations of the inner jet based on two models (upper right), restored with the same beam as the original map (lower right). Higher observing frequencies and space baselines are required to tell these apart. Figure from \citet{2019MNRAS.486.2873C}.}
    \label{fig:M87-jetsim}
\end{figure}

\begin{figure}[ht]
    \centering
    \includegraphics[width=0.3\textwidth]{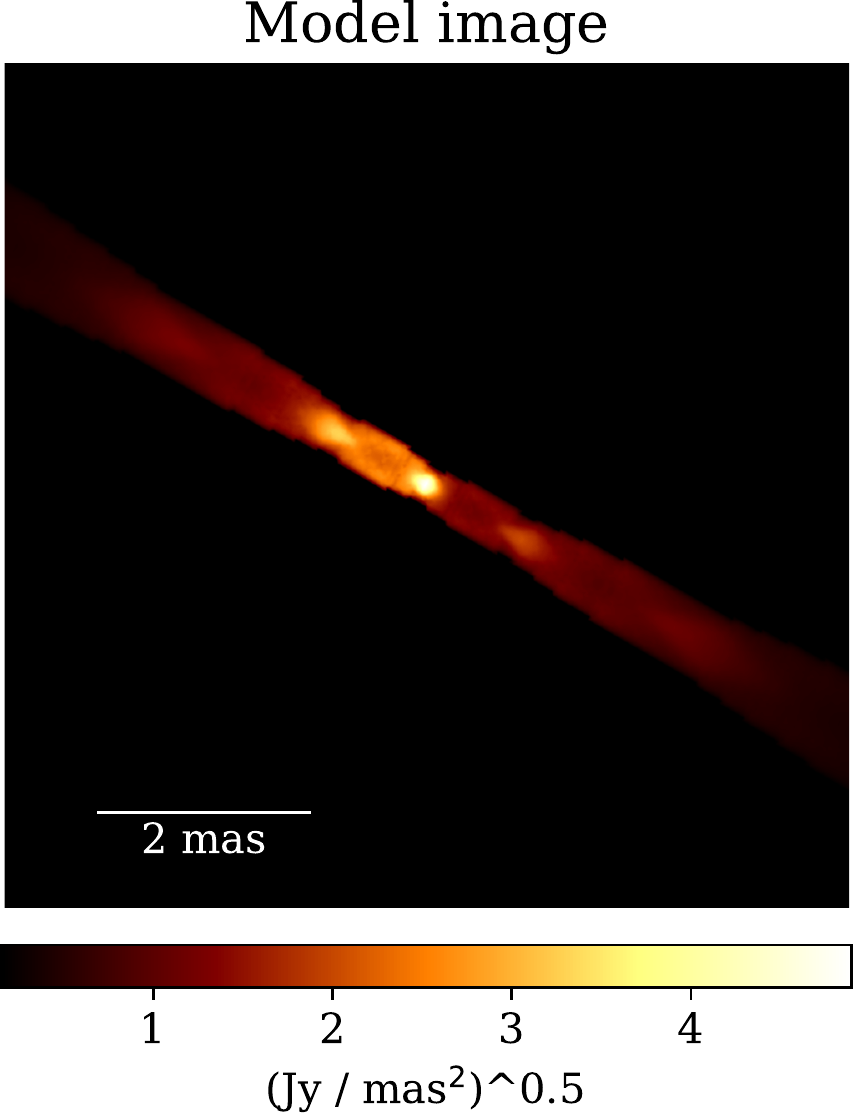}
        \includegraphics[width=0.3\textwidth]{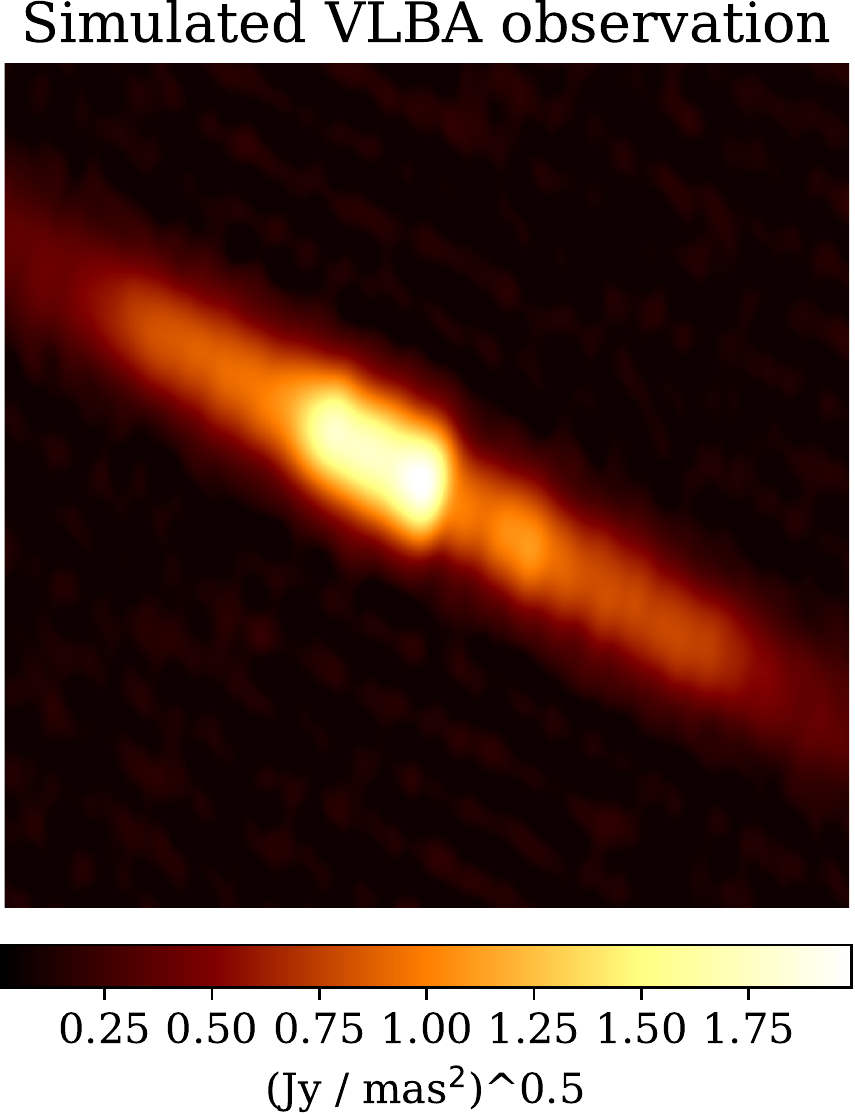}
            \includegraphics[width=0.3\textwidth]{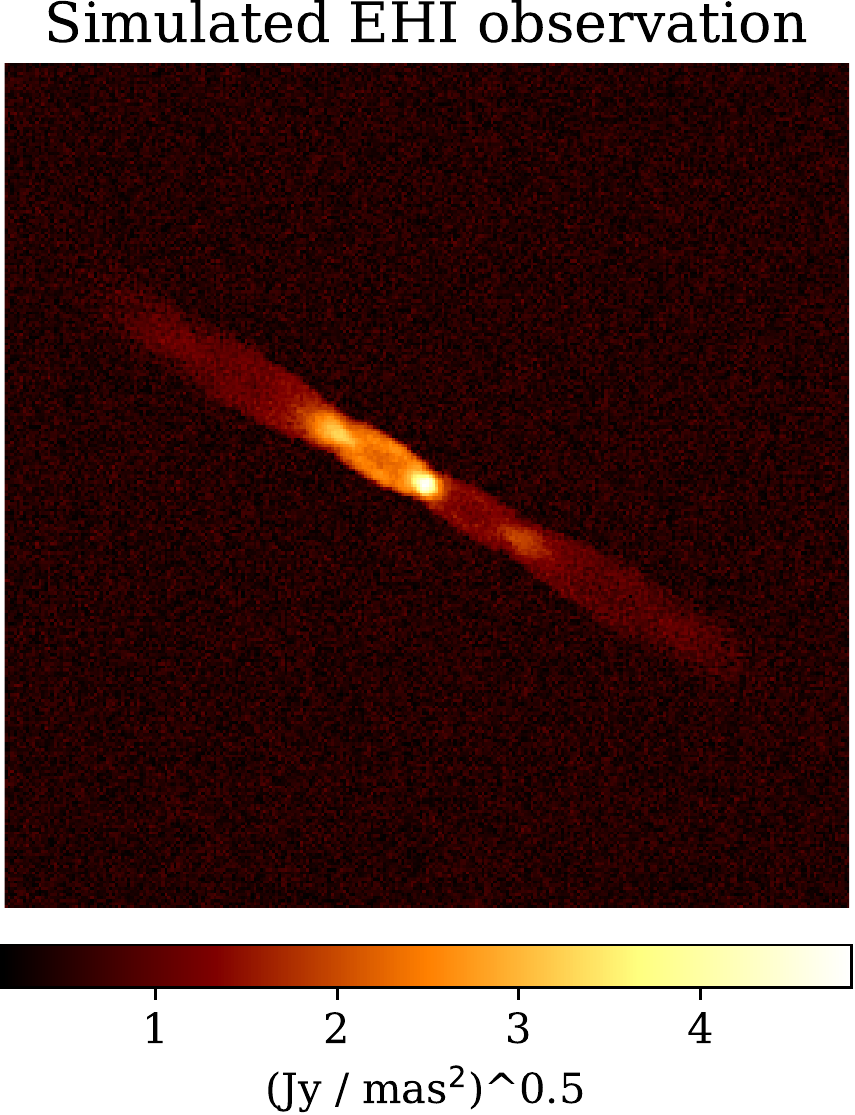}
    \caption{\textit{Left:} 43~GHz jet model with a recollimation shock \citep{Fromm2016,Fromm2018,Fromm2019}. The southwestern part is partly obscured by a dusty torus. The model is based on NGC1052 but can be used as a generic model for AGN jets. It was scaled to 8~Jy here to represent a typical bright AGN. Middle: simulated ground-based VLBA observation of this model. \textit{Right:} simulated Space VLBI observation of this model with the Event Horizon Imager (EHI, see Sec. \ref{sec:missionprofiles}).}
    \label{fig:frommjet}
\end{figure}

\subsection{Physics of inner jets in AGN}

The processes that govern the formation, acceleration, and collimation of powerful relativistic jets in active galactic nuclei (AGN) and X-ray binaries are a half-century old mystery in black hole physics. Without an instrument capable of resolving the accretion flow and eventual ejection of plasma in the immediate vicinity of the central black hole, most of the recent advances in our understanding of jet formation have resulted from general relativistic magnetohydrodynamic (GRMHD) simulations, e.g., \citet{McKinneyBlandford2009,TchekhovskoyNarayanMcKinney2011,Liska2018}. These simulations show that relativistic jets can be powered by magnetic extraction of rotational energy of either the central black hole itself, as originally proposed by \citet{Blandford1977}, or the accretion disc \citep{Blandford1982}. However, we still do not understand the details of this process and cannot answer such basic questions as how the properties of the accretion flow and black hole are connected to the jet formation and why only a fraction of the actively accreting supermassive black holes produces powerful jets in the first place.

In these magnetically driven scenarios, the jet is initially triggered by magneto-centrifugal force, with further acceleration and collimation produced by magnetic pressure gradients and tension forces. In these regions the jet is expected to be characterised by a parabolic collimation profile and a gradual transition from a predominantly poloidal to a helical or toroidal magnetic field configuration. While the jet launching takes place in the innermost few Schwarzschild radii, it appears that the collimation and acceleration extends up to $\sim 10^{5\pm1}$ Schwarzschild radii from the black hole \citep{Asada2012,Homan2015,Kovalev2019}, with a bulk of acceleration taking place within the first $\sim 10^{3}$ Schwarzschild radii \citep{Mertens2016}. To test jet formation models with actual observations it is necessary to probe linear scales smaller than this. With the exception of M\,87, the angular resolution required to resolve these structures is of the order of ten microarcseconds or better, and observations at short mm and sub-mm wavelengths are required to see through the self-absorbed synchrotron-emitting plasma at the jet base. This calls for Space VLBI with orbiting antennas operating at mm and sub-mm wavelengths.  

Space VLBI can provide the necessary angular resolution to study jet formation, collimation, and acceleration in other nearby AGN by directly resolving the sites of these physical processes \citep{Meier2009}. This would allow us to study what determines the jet power, and how this is related to the black hole spin, rate of accretion, and disc magnetisation. Especially, polarisation observations using Space VLBI are fundamental to reconstruct the three-dimensional magnetic field structure of the jet as near as possible to the black hole, thus helping to understand the jet formation, conversion of the magnetic energy to the kinetic energy of the flow, and dissipation of this energy. The likely co-existence of relativistic, projection and other effects makes it challenging to reconstruct the intrinsic orientation and strength of the magnetic fields along the jet where the very nature of these fields is still highly debated, see \citet{2017A-ARv..25....4B} and references therein. In particular, it is unclear whether the observed polarisation is due to compression in a shock of a random ambient magnetic field, or to the presence of a large-scale, ordered field permeating the plasma flow. Space VLBI polarisation imaging, both along and transverse to the jet direction, at millimetre wavelengths of a significant number of jets will enable us to reduce many of these uncertainties distinguishing between the different possible magnetic field configurations proposed, e.g., \citet{2016ApJ...817...96G,2008Natur.452..966M,2010ApJ...710L.126M}.

Furthermore, multi-frequency Space VLBI observations would allow measurement of a spatially resolved polarisation spectrum near the black hole, which can be used to probe the line-of-sight component of the magnetic field and the low energy end of the electron energy distribution, e.g., \citet{Homan2009}. Both of these are difficult to constrain by other ways. 

To illustrate the jet imaging possibilities with Space VLBI, we consider the jet model by \citet{Fromm2016,Fromm2018,Fromm2019}. VLBI observations of radio galaxies reveal highly collimated jets surrounded by obscuring dusty tori. Detailed studies of the jet structure revealed several regions of enhanced emission. These regions could be interpreted as travelling shock waves compressing the underlying jet flow or as stationary recollimation shocks. Recollimation shocks are formed due to the mismatch between the pressure in the jet and in the surrounding medium. They cause a distinctive radiative signature in the form of an edge-brightened flow converging into a central bright region, as seen between the two brightest spots in the model image in Fig.~\ref{fig:frommjet}. The recollimation shock profile cannot be resolved by the Earth-based VLBA, but a simulated observation with the two-satellite Event Horizon Imager Space VLBI concept (see Section \ref{sec:missionprofiles} and \citet{Martin2017,Kudriashov2019,RoelofsFalckeBrinkerink2019a} for details) shows that these features can be resolved with Space VLBI.

Combining Space VLBI with multi-wavelength observations that recover the complete spectral energy distribution (SED) from radio to gamma-rays will shed light on the mechanisms of energy dissipation in the innermost part of the jet, which is responsible for the spectacular flaring observed in blazars, e.g., \citet{2008Natur.452..966M,2010ApJ...710L.126M}. Correlated monitoring campaigns will help to localise emission features that fall outside the radio regime as we see structural changes related to flares at other wavelengths.

\subsection{Binary AGN -- Gravitational Wave precursors}

The early growth of massive black holes is believed to go through intense phases of accretion and mergers, and to be tightly coupled to the evolution of their host galaxies \citep{Hopkins+2008,KormendyHo2013}. Since mergers drive both gas and the black holes to the nucleus, this should often lead to pairs of active galactic nuclei (AGN). On kpc scales these can be easily resolved from the radio to the X-rays, but well-established dual-AGN are still rare \citep{2016IAUS..312...13K}. Direct imaging of such dual-AGN on parsec to tens of parsec scales requires milliarcsecond angular resolution and therefore can only be done using the very long baseline technique (VLBI) in the radio.  This means that a multi-band approach to confirm candidates is becoming problematic \citep{2018RaSc...53.1211A}. The best established case with a projected separation of 7~pc is 0402$+$379 \citep{2006ApJ...646...49R}. 

Below about 10 parsec separation the dual black hole system becomes gravitationally bound, and we refer to these systems as massive black hole binaries (MBHB), e.g., \citet{2015ASSP...40..103B} and references therein.  Over 100 MBHB candidates have been identified in optical quasar variability surveys \citep{Graham+2015,2016MNRAS.463.2145C,Liu+2019}, and orders of magnitude more is expected to be found in the first 5 years of operations of the much deeper all-sky time-domain survey to be conducted by the Vera Rubin Observatory in the framework of the Legacy Survey of Space and Time (LSST). It is estimated that as much as 1\% of the AGN below $z\sim 0.6$ could reside in massive binaries \citep{2019MNRAS.485.1579K}. This regime is particularly important because the processes that lead to hardening of the binary are not well understood. The energy loss in mergers of black holes is initially dominated by dynamical friction on stars and dark matter, and subsequently by stellar scattering, but below about a few parsecs this becomes inefficient. In this sub-parsec regime, gas may play a role in promoting binary inspiral. 

Somewhere between $100-1000$ gravitational radii the dominant energy loss becomes gravitational wave radiation and MBHBs become visible to LISA\footnote{www.lisamission.org, LISA site accessed 2020.12.24.}. Revealing the rates at which various processes work at intermediate scales is possible in principle, given a large sample of MBHBs, via a measurement of the relative abundances of MBHBs for a range of separations \citep{2009ApJ...700.1952H}. 

\begin{figure}[htbp]
    \centering
\includegraphics[width=0.42\textwidth]{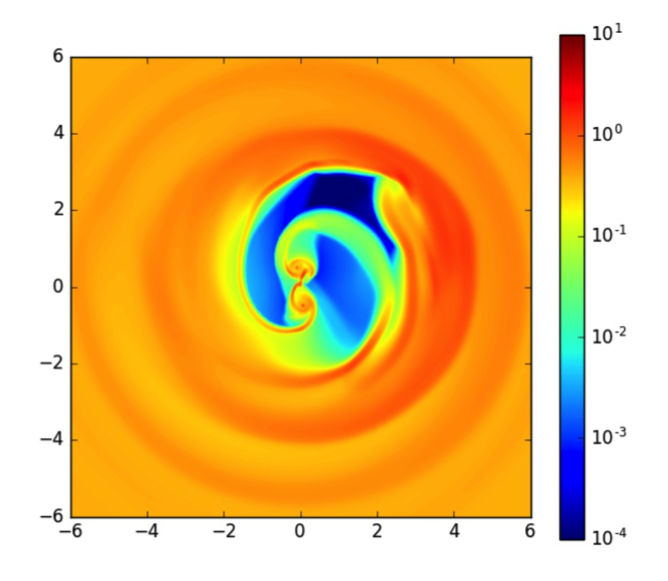}
\includegraphics[width=0.56\textwidth]{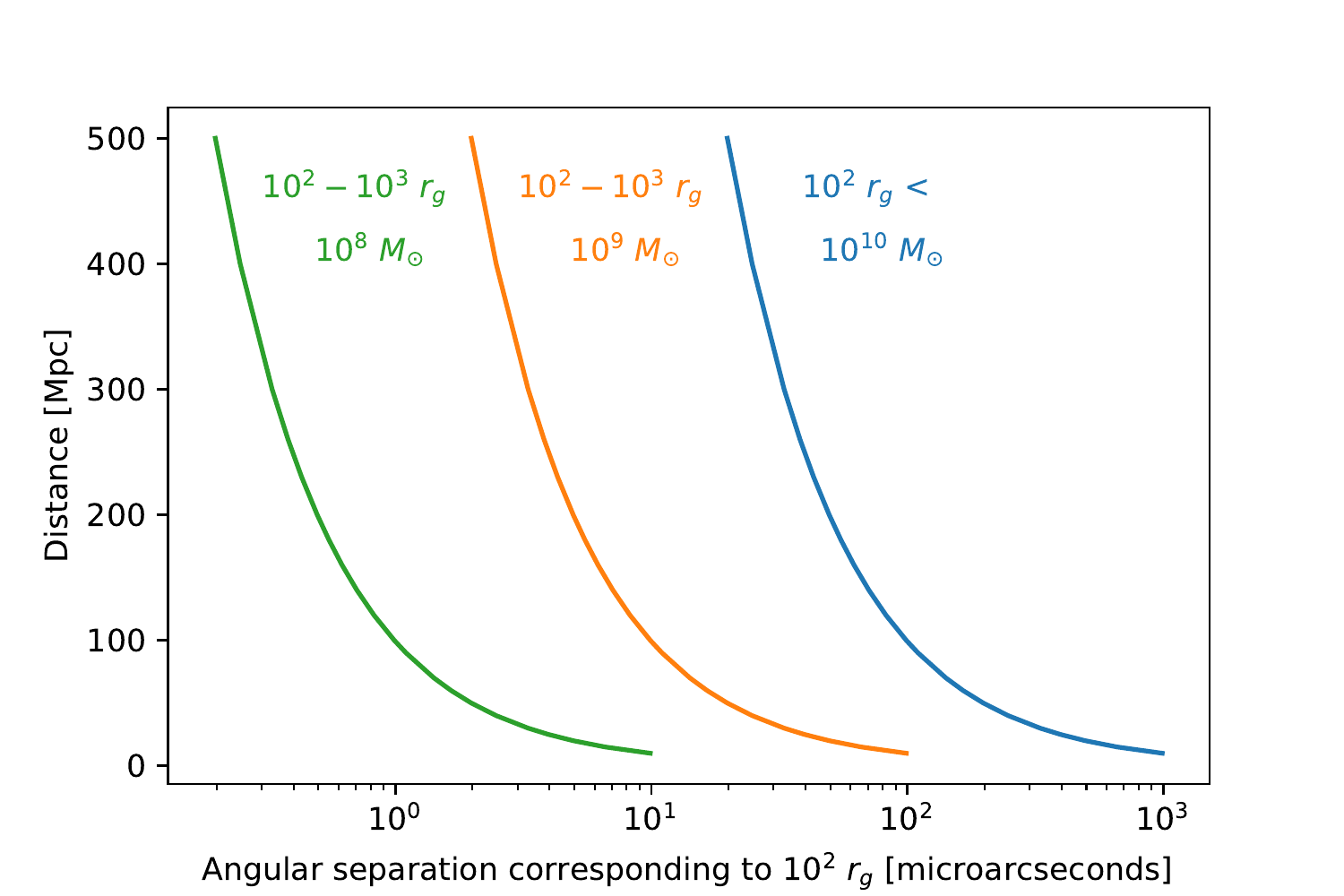}
    \caption{\textit{Left:} the surface density distribution in a sub-pc circumbinary gas disc \citep{2017MNRAS.469.4258T}, showing that tidal forces truncate mini discs in MBHBs, allowing the search for electromagnetic signatures of the binary. \textit{Right:} apparent angular size of $100-1000$ gravitational radii separation -- where GW radiation gradually becomes the dominant process in hardening the binary -- for various SMBH masses, in the local Universe. Binaries are expected to be long-lived (spending $\sim$100,000 years) in this regime \citep{2009ApJ...700.1952H} and can be directly probed by mm-VLBI for masses $10^8-10^{10}$~$M_{\odot}$. The most massive sub-pc separation binaries as well as jet formation in LISA-detected SMBH mergers will be observable at cosmological distances by THEZA (see also Fig.~\ref{fig:spin}).}
    \label{fig:SMBH-binaries}
\end{figure}

The possibility of directly resolving MBHB with mm-VLBI imaging was first discussed by \citet{2018ApJ...863..185D}. They focused on MBHBs that have orbital periods less than 10 years and are resolvable by ground-only baselines. They predict that ~100 such systems might be detectable down to 1~mJy limit within $z<0.5$ (their selection criteria for the tightest orbits and low Eddington rates for accretion biases towards the low-luminosity population). Studying the few hundred to few thousand gravitational radii regime for the most massive AGN ($\sim 10\%$ of which are radio-loud) at very high redshifts becomes feasible from space; estimates for the fraction of MBHBs in the AGN population range from $5\%$ \citep{2018ApJ...863..185D} to $30\%$ \citep{2015MNRAS.452.2540D,2016MNRAS.463.2145C}, Fig.~\ref{fig:SMBH-binaries}.

The caveat is that space mm-VLBI observations require that both black holes are active, which is quite rare in SMBH pairs with kpc separations. There are good reasons to believe this is not the case for sub-pc MBHB. This is because the most common MBHB are likely formed in minor mergers that have unequal mass black holes; below about 0.1~pc the binary is embedded in an accretion disc and gas interaction becomes important \citep{2002ApJ...567L...9A}. Hydrodynamic simulations indicate that in unequal-mass binaries the secondary gets most of the accretion, and starves the primary \citep{2015MNRAS.447L..80F}. This will likely make the secondary super-Eddington, and the primary sub-Eddington (aka \lq\lq hard state'') -- both of these \Mdot\, regimes are associated with jets and radio emission.

Depending on the configuration, a space mm-VLBI array operating at 230~GHz would have an angular resolution from 10 $\mu$as (few Earth radii) to 1~$\mu$as ($\sim$25 Earth radii) allowing direct imaging of a number of candidates. For reference, the diameter of the estimated orbit of the $z=0.3$ periodic binary candidate quasar PG\,1302$-$102 extends $\approx$10~$\mu$as \citep{2015Natur.518...74G}. 

\subsection{Time- and X-ray-domain synergies}

\begin{figure}[t]
\centering
\includegraphics[width=0.99\textwidth]{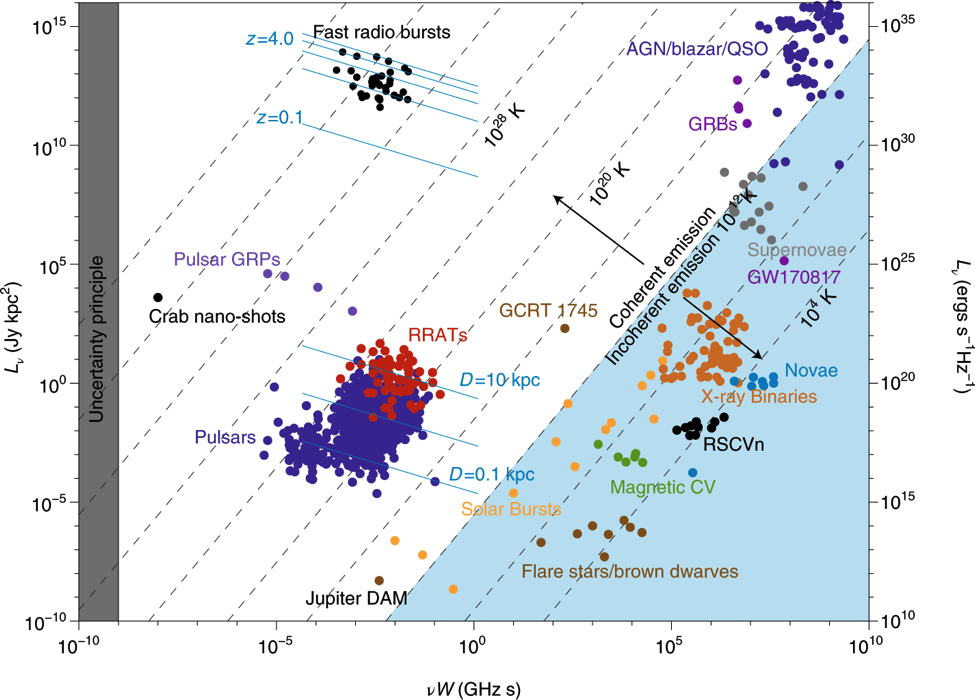}
\caption{Luminosity -- timescale parameter space for radio transient and variables. The blue shaded region delimits a brightness temperature of $10^{12}$K -- sources inside this region are likely to be incoherent synchrotron emitters, those in the white region coherent. From \citet{2018NatAs...2..865K} and \citet{2015MNRAS.446.3687P}.}
\label{fig:transients}
\end{figure}

Radio transients are both the sites and signatures of the most extreme phenomena in our Universe: e.g. exploding stars, compact object mergers, black holes and ultra-relativistic flows. Essentially all explosive events in astrophysics are associated with incoherent synchrotron emission, resulting from ejections at velocities in excess of the local sound speed that compress ambient magnetic fields and accelerate particles. These events range from relatively low-luminosity flares from stars, to the most powerful events in the Universe, associated with gamma-ray bursts and relativistic jets from supermassive black holes in active galactic nuclei. Crucially, radio observations can act as a calorimeter for the kinetic feedback, probe the circumburst environment, and provide localisation/resolution unachievable at other wavelengths. Follow-up radio observations of high-energy astrophysical transients has a rich history of important discoveries, including the first galactic superluminal source \citep{1994Natur.371...46M}, the beamed-like nature of Gamma-Ray Bursts (GRBs) and their association with unusual supernovae \citep{1998Natur.395..663K} and the association of highly relativistic jet-like flows with the tidal disruption and accretion of a star by a supermassive black hole \citep{2011Natur.476..425Z}. Most recently, the LIGO-Virgo binary neutron star merger GW170817 was associated with a relativistic jet and radio afterglow \citep{2018Natur.561..355M,2019Sci...363..968G}. Fig.~\ref{fig:transients} illustrates the luminosity -- timescale space for all radio transients, including coherent sources (see below).

\begin{figure}[!ht]
\centering
\includegraphics[width=0.8\textwidth]{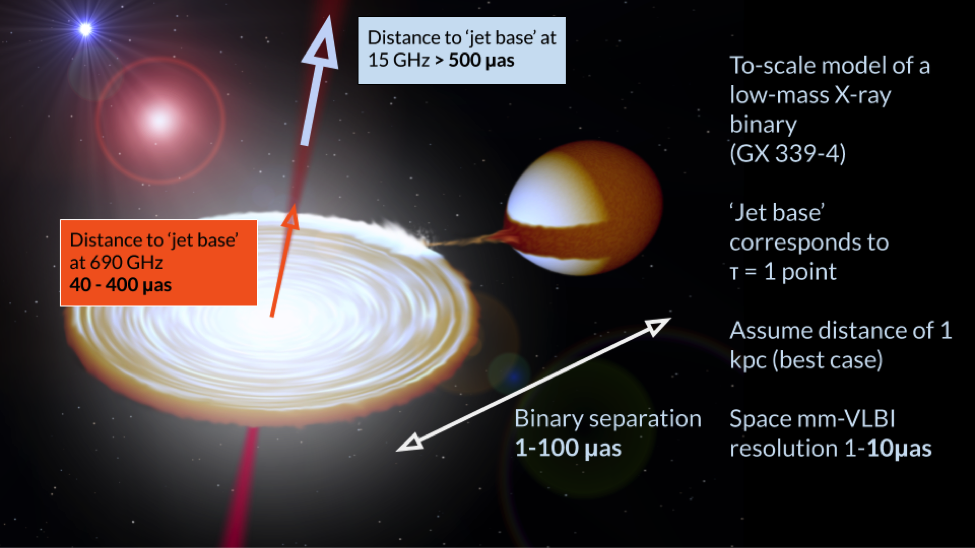}
\caption{Illustration of the angular scales associated with a low-mass black hole binary system at a distance of 1~kpc (equal to the smallest current distance known, typical distances are 8~kpc). For a space-VLBI resolution of 4~$\mu$as at 690 GHz both the jet base and binary separation are clearly resolvable. Coupled with X-ray (inner accretion flow) and infrared (innermost regions of the jet, close to launch zone) observations, we could achieve unprecedented understanding of how jets form and couple to accretion.}
\label{fig:bhxrb}
\end{figure}

A prominent example of transient radio emission is accretion onto black holes, in which one of the most relativistic and energetic processes in the Universe occurs on a range of spatial and time scales which extends over more than seven orders of magnitude. On the smallest scale is accretion on to stellar mass black holes in binary systems, in which we can track the full evolution of the radio jet and its connection to the varying accretion flow  humanly-accessible timescales. Since these systems are typically at the same distance as Sgr~A* (the closest known is about a factor of 8 closer), but they are 100,000 times less massive, we will never be able to image on scales comparable to their event horizons. However, we are able to test fundamental regions of the inner relativistic jet and, given the unique empirical coupling with the accretion flow we have established in such systems, probe the physics of time-variable jet formation in a way not really possible for AGN. Fig.~\ref{fig:bhxrb} illustrates the spatial scales associated with a typical, nearby, black hole binary in our galaxy and shows how the innermost regions of the jet could be imaged by space-based mm VLBI. At these scales we would expect to see variations in the $\sim$ THz jet base occurring only {\em minutes} after X-ray variations, providing clues to the formation of jets in unprecedented detail. On larger physical scales, highly transient radio emission has been clearly associated with the transient accretion onto massive black holes resulting from tidal disruption of a star. In a non-thermal Tidal Disruption Event (TDE), this emission is likely to arise in a jet in most, if not all, cases, and VLBI offers a unique opportunity to see how accretion and jets evolve in pristine environments rather than in very-long-timescale steady states associated with most AGN; see \citet{2016MNRAS.462L..66Y,2018Sci...361..482M}. 

At the most explosive end of the stellar scale are gamma-ray bursts (GRB). The long variant is associated with the death of the most massive stars, while the short variant now appears quite clearly to be associated with the merger of two neutron stars. The end product in both cases can be stellar-mass black holes accreting at highly super-Eddington rates, and these are known to be powerful engines of relativistic jets. For a long time, long GRBs have been known to produce relativistic jets that can be studied by ground-based VLBI \citep{2004ApJ...609L...1T}. Recently superluminal motion was detected in a jet associated with the late-peaking ($\sim 6$ months) radio afterglow of the neutron star merger GW170817 \citep{2018Natur.561..355M,2019Sci...363..968G}. In exceptional cases, we might be able to study the ultra-relativistic phase in \lq\lq engine-driven'' stellar jets by Space VLBI, if these are nearby and beamed in our line of sight. 

Coherent transients and variables are amongst the most important sources in astrophysics right now, in particular Fast Radio Bursts (FRBs) and pulsars. Often the broad-band nature of the phenomena is not well established (e.g. for FRB). Prominent counter-examples include pulsars and neutron stars, which are observed across the whole electromagnetic spectrum, but even in those cases, the THz window is mostly unexplored. Potential sources for this frequency range are, for instance, magnetars, which have been detected up to 291~GHz \citep{2017MNRAS.465..242T}  with rather flat flux density spectra spanning from a 1~GHz up to, now,  300~GHz. Whether this makes them prominent sources at THz frequencies will be interesting to see. The current status of the transient sky at radio frequencies is summarised in Fig.~\ref{fig:transients} as compiled in \citet{2018NatAs...2..865K}. It is likely that THz frequencies will mostly probe the incoherent part 
but the magnetar detections at 300~GHz suggests that we should be open for surprises.

\subsection{Water maser science with Space VLBI}

Water vapour is found out to cosmic distances, e.g., \citet{2008Natur.456..927I}. It can emit thermally or through stimulated emission, i.e., masers. %Masers are beacons of their pumping mechanism, collisions or radiative excitation, that reflects particle densities and temperatures. 
Since masers are compact and bright, they are excellent astrometry targets. Due to heavy atmospheric absorption the mm and sub-mm masers are much harder to observe from Earth than the well-studied 22~GHz transition. While bright water masers, at, e.g., 183~GHz and 325~GHz, can be observed by dedicated ground-based telescopes (see \citet{2017A&A...603A..77H}), others have only been detected from above the main atmosphere layer by the Kuiper Airborne Observatory and SOFIA, e.g.,  \citet{2017A&A...606A..52H}; \citet{2017ApJ...843...94N}; \citet{2019inpressRichards} and references therein. 
%Unlike the 22\,GHz water maser, some of the mm-range water masers (e.g. emitted at 183 and 325\,GHz) are not easily observed from Earth due to strong atmospheric absorption (e.g. %Richards et al. 2014, 
%\citealt{2017A&A...603A..77H}). %Water masers in the mm are generally believed to be excited by a similar pumping mechanism to 22\,GHz masers (\citealt{2013ApJ...768L..38H}). 
There are over a hundred predicted water maser transitions from GHz--THz frequencies (a few tens have been detected), mostly excited by collisional pumping under distinctive combinations of temperature, number density and other parameters \citep{2016MNRAS.456..374G}. ALMA and RadioAstron observations found that mm and 22\,GHz water maser sources can have both extended and extremely compact morphologies, down to just a few solar radii \citep{2012ApJ...757L...1H,2018IAUS..336..417S}. The millimetre water transitions are thus very suitable for space-space VLBI that would cover a range of baselines from hundreds of km to tens of Earth radii.
%though some transitions may be pumped in colder and less dense environments, and are thus expected to have more extended but also extremely compact,  few Solar radii) emission, as was observed for 22 GHz on satellite to ground-based telescope baselines (e.g., Sobolev et al. 2018).

\begin{figure}[!ht]
\centering
\includegraphics[width=0.65\textwidth]{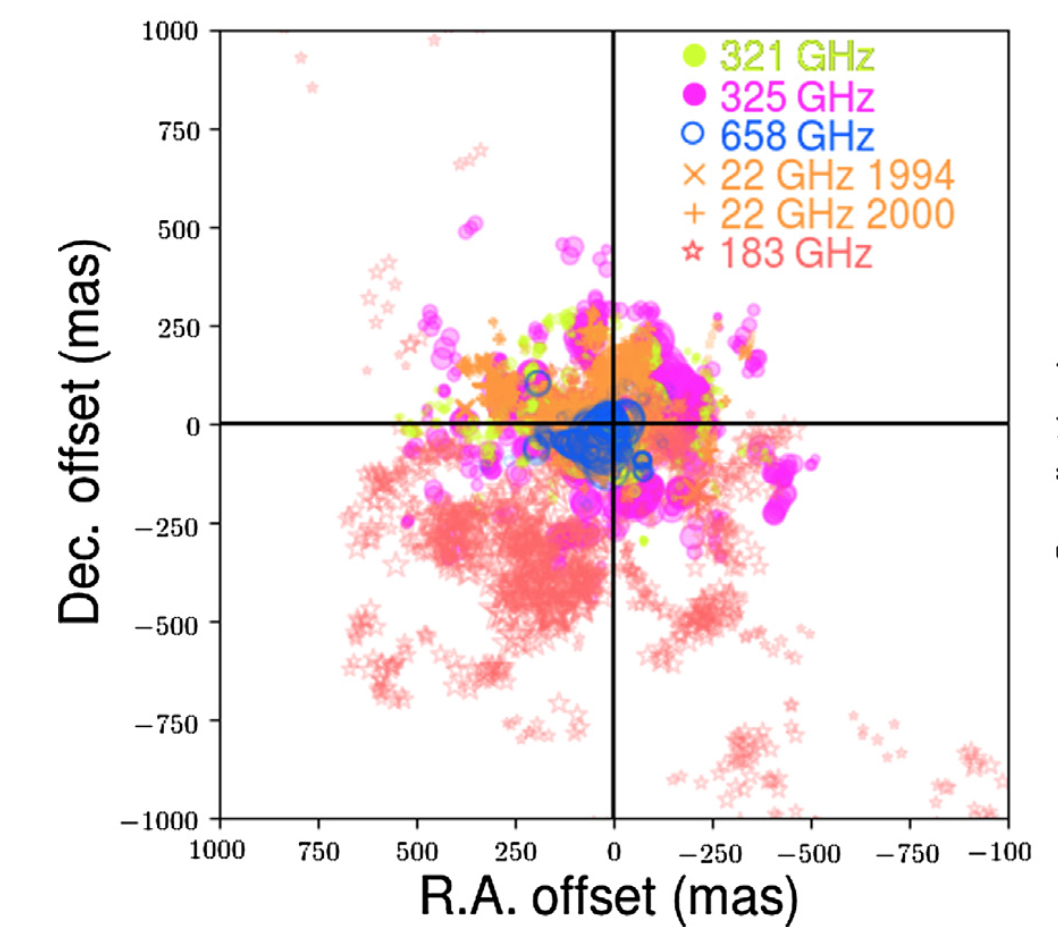}
\includegraphics[width=0.6\textwidth]{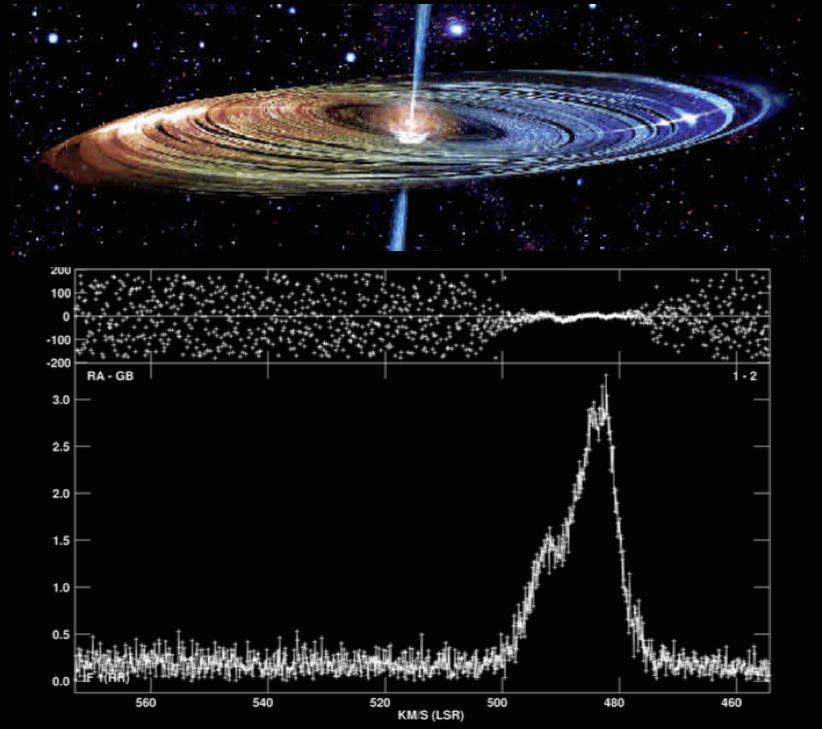}
\caption{\textit{Top:} Various water maser transitions in the envelope of VY CMa \citep{2019inpressRichards}. \textit{Bottom:} Water masers in the circumnuclear disc of NGC4258 (artist impression: J.\, Kagoya, M.\,Inoue): RadioAstron satellite to Green Bank Telescope baseline of 1.9~Earth diameters reveals the presence of compact, 790~AU, masing clouds \citep{2018IAUS..336..422B}.} %H$_2$O maser detection on a 25 Earth diameter baseline in the galaxy NGC~4258 with the RadioAstron satellite (Baan et al. 2018).}
\label{fig:masers}
\end{figure}

Typically, Galactic water masers arise from a disc--jet system of a young stellar object and in envelopes around evolved stars. AU-scale spatial distributions of a number of mm water maser transitions %combination with the 22\,GHz water (n, T) and methanol masers (n, T) would help us 
can put strong constraints on density, temperature and water distribution in gas surrounding the young stellar object that could be compared with current theories of star-formation, e.g., \citet{2016ApJ...823...28K}; they provide powerful diagnostics on the disc chemistry \citep{2015ApJ...815L..15B}. Proper motions can trace the dynamics of the gas flow, and reveal the disc or jet origin of the emitting gas, as these have distinct kinematic signatures \citep{2015A&A...583L...3S}. %, for which xx au-scale temperature and density probes are required. 
For evolved stars, various water transitions are observed at different stellar radii, see Fig.~\ref{fig:masers},  tracing the change of physical conditions in the stellar envelope \citep{2014A&A...572L...9R}. 

For nearby and distant galaxies, bright mm water maser emission is an excellent tracer for both the star formation processes and physics of the activity of their central engines, in some cases -- AGN. For example, in the interacting galaxy Arp 220, ALMA observations of mm-water masers find unresolved emission that is best modelled by a large number of pc-scale molecular clouds \citep{2017A&A...602A..42K} -- only space--space VLBI could provide images at these scales. On the other hand, towards Circinus and NGC\,4945, the mm water maser emission is associated with the circumnuclear region \citep{2013ApJ...768L..38H,2016ApJ...827...68P} . Water masers from circumnuclear discs have been found to have extremely compact components on Space--Earth baselines, see Fig.~\ref{fig:masers}. (Sub)mm water maser emission from the circumnuclear discs surrounding supermassive black holes holds the promise of yielding independent estimates of geometric distances, and perhaps even an independent verification of the Hubble constant.

One of the major benefits of space--space VLBI is the absence of the terrestrial atmosphere: (sub)mm water masers are currently difficult or even impossible to observe at most ground-based sites (see \citet{2016MNRAS.456..374G} for model predictions of millimetre maser transitions). SOFIA observations have already detected THz water maser emission at 1.3~THz \citep{2017A&A...606A..52H}. The more transitions of water that can be observed from the same volumes of gas, ideally quasi-simultaneously, the better constrained radiative transfer modelling to determine density and temperature become. Detecting the lines for the first time also provides a good test of the predictions of maser theory and models.

\subsection{The quest for water in protoplanetary discs}\label{Sect:Water-discs}

Water is an important molecule in many astrophysical environments. In the case of protoplanetary discs, two key issues emerge from previous studies. On the one hand, water can freeze out on dust grains if conditions get sufficiently cold and shielded. In the heavily discussed core accretion scenarios of planet formation, the presence and location of such ice lines are essential ingredients for determining where and how efficiently the small dust grains can stick together and start the growth to larger agglomerates as a first step to planet formation, e.g., \citet{2016SSRv..205...41B}. Analysing the water vapour signals from the gas phase gives crucial constraints on the distribution of the water phases in a disc.

Secondly, water is intimately linked to the composition of exoplanets, e.g., \citet{2019BAAS...51c.229P} and references therein. During the process of planet formation, the abundance and phase (solid or gas) of water traces the flow of volatile elements with implications for the bulk constitution of the planets, the composition of their early atmospheres, and the ultimate incorporation of such material into potential biospheres, e.g., \citet{2012E-PSL.313...56M,2017PNAS..11411327P}. Furthermore, water as a simple molecule with high abundance is the dominant carrier of oxygen. Hence, its distribution in a disc can also steer the C/O ratio which can be a tell-tale connection between planet composition and the natal disc composition and location of the birth site (especially for giant planets).

After all, the complicated interplay between grain evolution, grain surface chemistry and freeze-out, photodesorption and photodissociation, and radial and vertical mixing processes will regulate the abundance of water in its different phases especially in the outer disc, e.g., \citet{2013ChRv..113.9016H}. But integral models of these processes can only advance if we have observational access to all phases of water at all temperature regimes in a protoplanetary disc. Here, we will still have major deficits even in the 2030s.

Though water vapour lines have been detected from the ground, such observations are mostly limited to high-excitation thermal lines ($E_{\rm up}/k \gtrsim 700$~K) that are not excited in the Earth's atmosphere, or to certain maser lines. Water lines seen in the near-infrared just arise from the very inner hot gas disc. In the mid-infrared, one still probes very warm water gas of many hundred kelvin. The MIRI instrument on the James Webb Space Telescope (JWST) will make very sensitive observations of such warm water lines for many protoplanetary discs. But the spatial resolution of JWST in the mid-infrared towards longer wavelengths is limited, and typical discs of just 1--4 arcsec in size will just be moderately resolved. Furthermore, JWST does not offer sufficient spectral resolution to resolve the velocity structure of the detected lines. To get access to the bulk of the water vapour reservoir in a disc that contains the colder gas, one needs to include the far-infrared and sub-millimetre range. With the aforementioned difficulties for sensitive observations of cooler water vapour from ground that severely hamper even ALMA with its good observing site, observations from space are pivotal to make progress in this field. To get spatially resolved information on such lines in a typical disc, one should aspire to an angular resolution of 0.1''. Thus, this field of research would demand short baselines on the order of several hundred metres to a kilometre.

\subsection{Exoplanets}

Thousands of exoplanets have been discovered in the last decades, showing the ubiquitous character of the planetary objects. The discovery of new worlds will continue in the next years even more efficiently with new and more sophisticated instruments, which will help to provide a complete knowledge of the formation and evolution processes of these objects. Observations are being carried out covering the complete electromagnetic spectrum from visible, infrared to radio wavelengths. However, and despite the extraordinary contribution of the ALMA interferometer to protoplanetary discs, the observation of exoplanets at millimetre and sub-millimetre wavelengths, covering to THz frequencies, is yet to be developed. Ground-based THz astronomy is heavily hampered by absorption features in the Earth’s atmosphere, caused by the presence of molecules of oxygen and water. Although Earth’s atmosphere's opacity can partially be alleviated by ground observatories built at high altitude, only space-based interferometers would definitely be free from Earth’s atmospheric limitations, fully exploiting the science provided by sub-mm wavelengths.

Detection of exoplanets via astrometric monitoring of the reflex motion of the parent star can be successfully applied in the sub-mm range. The stellar photosphere is dominated by the permanent and more stable thermal, black body emission ($\propto\nu^2$), free from the on/off nature of stellar flares present in active stars at longer wavelengths. For the nearest stars, radio luminosities of normal, solar-like stars would correspond to tens of mJy at 100s GHz frequencies \citep{Lestrade2008}. An efficient monitoring of selected samples would solve the orbit inclination ambiguity inherent to the planetary masses determined by radial velocity. Additionally, the population of planets on the outer side of the planetary systems, less sensible to radial velocity techniques, can be characterised. Sub-milliarcsecond-precise astrometry would suffice to detect Jovian planets within 10~pc.

Direct detection of exoplanets at sub-mm wavelengths would require $\mu$Jy sensitivities. Most of the continuum exoplanet emission would correspond to thermal radiation, which at THz frequencies may reach 10-100 $\mu$Jy from a Jupiter-like planet within 10\,pc \citep{Villadsen+2014}. This emission is 1-2 orders of magnitude larger than that measured by ground-based telescopes, showing that direct detection of exoplanets will necessarily require space missions. The importance of these measurements is fundamental as they will be able to bridge the gap between the far- and mid-IR studies of exoplanets with the, so far elusive, emission at long radio wavelengths. Perspectives to detect exoplanets during their early stages of formation are particularly favoured at sub-mm wavelengths. Protoplanets still embedded in the circumstellar discs will radiate by reemission of the heated surrounding dust. Even modest millisecond resolution would discriminate between the emission of the disc and that of the circumplanetary material for a Jupiter at 1\,AU within 100\,pc. Measurements of non-Keplerian motions of the protoplanet would provide direct information of the density and viscosity of the disc \citep{Wolf+D'Ang2005,Pinte+2018}.

The sub-mm range of the spectrum (100--1000~GHz) is particularly rich in water lines which, with the adequate sensitivity and resolution, may not only characterise the atmosphere of newly discovered planets but unambiguously trace the signs of biological activity. Sub-mm spectroscopy of exoplanets highly irradiated by their host stars constitutes the best scenario to detect these absorption features. These studies are out of the capabilities of existing, or even planned, ground-based observatories as they are only doable for space-based missions with $\mu$Jy sensitivities \citep{Oberg+2018}. The characterisation of the atmosphere of Earth-like planets by high-resolution sub-mm observations is of extraordinary relevance as the presence of water is anthropologically linked to the development of life.

\subsection{SETI -- search for technosignatures}

Over the last few years, the field of SETI (Search for Extraterrestrial Intelligence) has undergone a major rejuvenation, see, e.g., \citet{Price+2019,Siemion+2013}. The discovery by the Kepler mission that most stars host planetary systems, and that around 20\% of these planets will be located within the traditional habitable zone, plus the continually growing evidence that the basic pre-biotic constituents and conditions we believe necessary for life are common and perhaps ubiquitous in the Galaxy, has brought new focus to one of the most important questions that human-kind can ask itself: Are We Alone? 

A space-based mm-VLBI interferometer operating outside of the Earth's atmosphere would enable the first serious SETI searches to be conducted across the full mm and sub-mm domains of the electromagnetic spectrum. The recent upsurge in interest in the search for ``techno-signatures''  \citep{Wright2019} has a strong focus on covering as much of the electromagnetic spectrum as is sensible.  Searches at mm and sub-mm wavelengths are extremely well motivated. In particular, advanced civilisations will be well aware of the advantages of operating communication systems within this particular ``high frequency" domain, especially for long distance communication systems that are likely to be associated with powerful interplanetary (and indeed interstellar) networks. Large bandwidth sub-mm systems offer significant carrying capacity, and yet continue to operate in a regime where scattering by ionised gas, and absorption by dust are usually negligible. This part of the high frequency radio spectrum is also relatively free of  human-made radio frequency interference. Although this almost pristine environment is likely to change in the coming decades, a space-based, sub-mm long-baseline interferometer is largely immune to the effects that plague ground-based arrays of limited spatial extent.

The detection of ``leakage" radiation or perhaps deliberately established beacons from another technical civilisation would be possible with a space-based sub-mm interferometer. In particular, such signals are likely to be entirely unresolved, even with resolutions approaching 10~$\mu$arcsecs. Narrow-band signals (like beacons) would probably exhibit very large Doppler accelerations and those transmitters associated with exoplanets located within 1~kpc of the Earth and with similar orbital periods ($\sim 1$~yr), would show changes in proper motion to be detected that a space-based sub-mm interferometer would be sensitive to on timescales as short as a few days. 

A space-based interferometer might also be able to detect non-coherent technosignatures. For example, the emission of waste heat from highly efficient mega-structures (such as Dyson spheres/swarms) could re-emerge as relatively cold black-body emission in the sub-mm. In particular, discontinuous structures such as sharp edges or holes (as might be commonly associated with artificial mega-structures in general) would induce ringing in the $uv$-plane and a highly correlated response in the visibility data. 

\section{THEZA implementation options}
\label{sec:missionprofiles}

The key capabilities of the THEZA concept require a space-borne VLBI system able to observe at frequencies above 200~GHz (1.5~mm wavelength) to at least 1~THz (300~$\mu$m wavelength) or even higher. Extension of the observing range toward lower frequencies, e.g., down to 86~GHz might be considered as an attractive broadening of the THEZA science scope. A design of the mission addressing the THEZA science outlook will be the subject of several major engineering trade-offs. One of them is between interferometers employing Space--Earth and Space--Space-only baselines. 

The former would have an enhanced baseline sensitivity due to a larger collecting area of Earth-based antennas, ultimately like a phased ALMA, just as demonstrated recently by \citet{EHT2019I}. However, such the system will be limited in frequency coverage by the atmosphere opacity thus likely operating efficiently at frequencies not higher than 350--400~GHz (practically at 230~GHz, due to a foreseeable lack of Earth-based facilities able to operate at frequencies above 230~GHz simultaneously). That said, the ongoing design studies of several Earth-Space mm-VLBI mission concepts, such as Millimetron \citep{Kardashev+2014} should provide useful input into assessment of the feasibility of various models of THEZA implementation.

Interferometers with Space-only baselines offer a clear advantage in frequency coverage: they are not subjected to severe atmosphere limitations and can operate at frequencies above the practical for Earth-based VLBI cut-off of around 300~GHz. The further advantage is a principle possibility to cover the $uv$-plane efficiently by using free-flying spacecraft formations. However, these advantages come at a price: it is unrealistic to expect that on the timescale of the Voyage 2050 programme, large mm/sub-mm space-borne apertures comparable in size to Earth-based antennas can be deployed. Yet, by using receivers with system temperatures near the quantum limit and wide-band data acquisition systems, an acceptable baseline sensitivity can be achieved even with moderate in size space-borne antennas. This approach is pursued in several ongoing studies, including IRASSI (Infrared Astronomy Satellite Swarm Interferometry) for far-infra-red astronomy \citep{Linz+2019} and a concept of Event Horizon Imager (EHI), a step beyond the EHT requiring space-borne interferometric elements \citep{Martin2017,Kudriashov2019,RoelofsFalckeBrinkerink2019a}. 

\subsection{Event Horizon Imager: a case study of space-only sub-mm aperture synthesis system}

\begin{figure}[ht]
\centering
\includegraphics[width=0.7\textwidth]{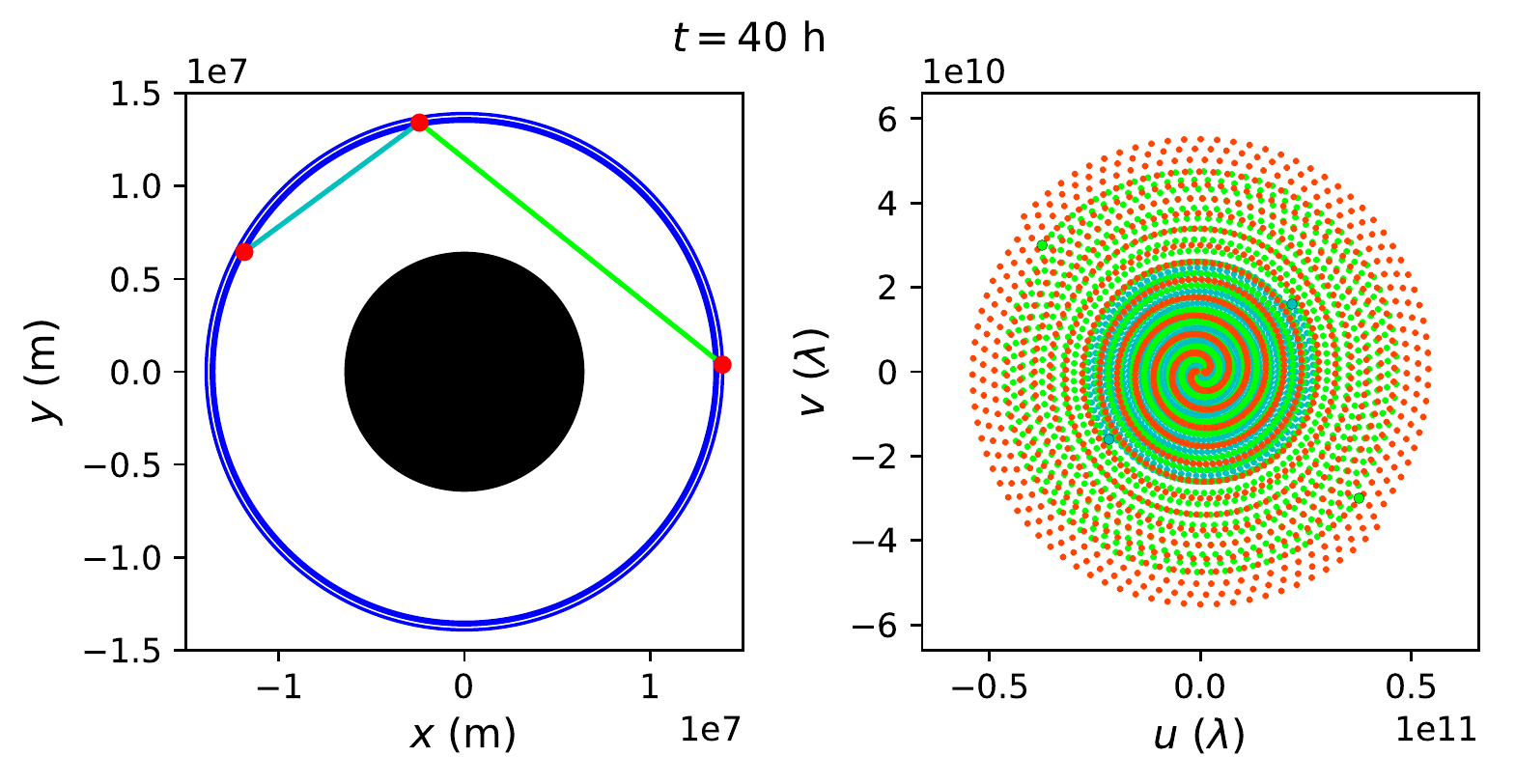}
\caption{\textit{Left:} Satellite orbits (blue) around Earth (black disc) with current satellite positions indicated in red and baselines indicated in green and blue. \textit{Right:} Corresponding $uv$-coverage. The maximum baseline length is set by the point where the Earth occults the intersatellite link between two satellites. Further details and figure credit:   \citet{RoelofsFalckeBrinkerink2019a}, reproduced with permission \copyright ESO.}
\label{fig:ehispiral}
\end{figure}

The EHI concept considers two or three satellites in circular medium-Earth orbits (MEOs). By setting a small difference between the orbit radii, the satellites drift apart as they orbit Earth, increasing the baseline length as it constantly changes orientation. The resulting $uv$-plane coverage will have the shape of dense and isotropic spirals, which is especially suitable for high-fidelity and high-resolution imaging (Fig.~\ref{fig:ehispiral}). In fact, the $uv$-coverage will be unlike that of any interferometer before and be almost like a filled aperture for integration times of weeks. Filling of the $uv$-plane is essentially done by Earth's gravity and happens in principle without any active orbital control during an observation. 

Adjustments of the orbital height separation allow one to adjust the configuration to fill the $uv$-plane within a desired time scale from a few days to months, commensurate with the variability or integration time scale of the source to be observed. For small orbital height separations even a compact configuration with only a few hundreds of meters to hundreds of kilometres could be achieved and maintained for an extended period of time (e.g., offering a smooth extension of the resolving power of ALMA). 

Otherwise the baselines will be significantly longer than any ground-baselines and shorter than some of the longest baselines in past Space VLBI experiments, but ideally matched to the desired resolution with maximum image fidelity. 

Space-space operation allows one to go to much higher frequencies than with Earth-based arrays or Space-Earth experiments, which increases resolution further. 

The concept aims to exchange the data between the satellites via a laser-based intersatellite link and correlate the data on-board using an orbit model provided by, e.g., measurements with GNSS satellites. Circular MEOs allow for a relatively stable orbit to start with. Further processing of the data is then done on the ground using a refined orbit model based on, e.g., intersatellite ranging measurements and astronomical calibrators. The local oscillator signals may be shared between the satellites as well in order to increase phase stability. An on-going technological study is investigating the feasibility of this concept \citep{Martin2017,Kudriashov2019}.

The laser-based intersatellite link provides high data rates over very large space-space distances and hence allows wide bandwidths and accordingly high sensitivity even with modest-sized dishes. The laser links could also be used to directly transport the IF signal down to Earth. That would allow one to perform Space-Earth VLBI during special campaigns, e.g. to do snap-shot observations of highly variable or very faint objects together with sensitive ground-arrays (e.g. EHT/ALMA or ngVLA). This is obviously only possible at longer wavelengths, e.g. 1.5~mm, as demonstrated by the EHT, or 3~mm, as done regularly by the Global mm-VLBI Array.  

The concept of multi-element space-borne interferometer has been also considered by \citet{Fish+2019}. Their concept differs from the EHI in the choice of orbits, number of space-borne antennas and the overall interferometric configuration of the system. These concepts have a lot in common and offer convenient starting points for further mission design studies. 

\subsection{Simulations of the Event Horizon Imager}

\begin{figure}[ht]
\centering
\includegraphics[scale=0.36]{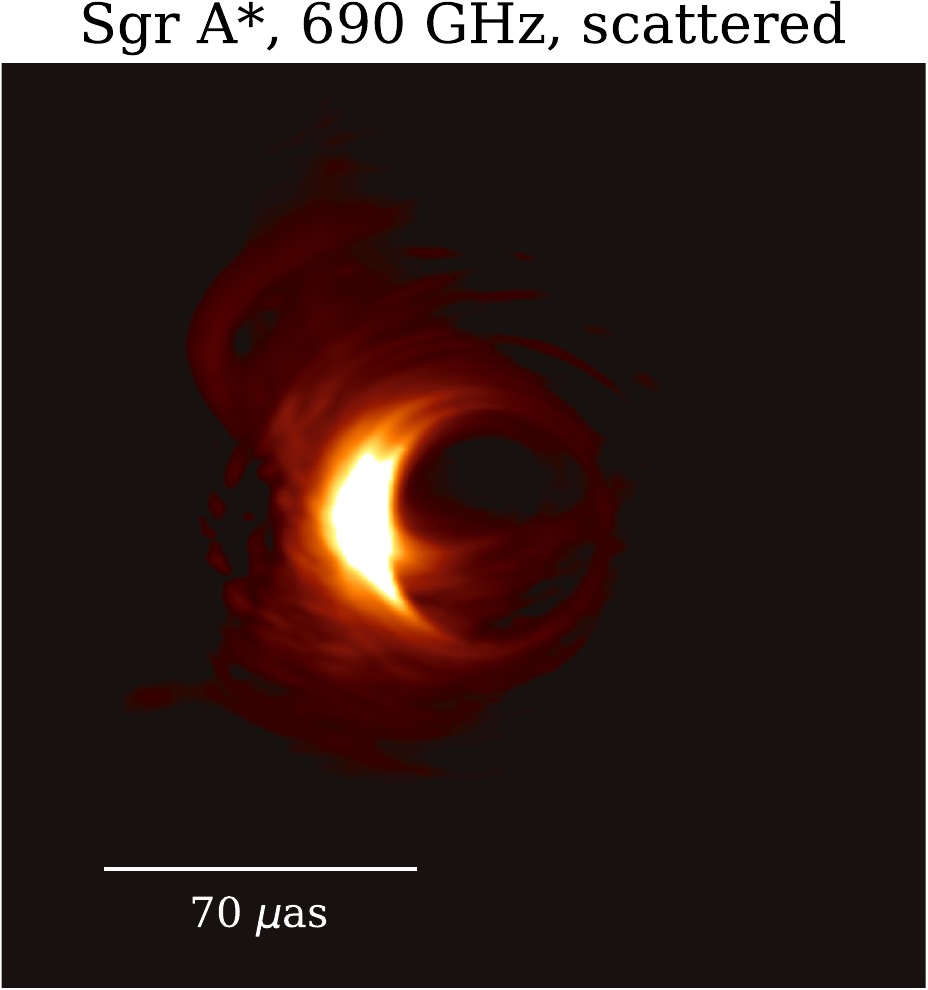}
\includegraphics[scale=0.36]{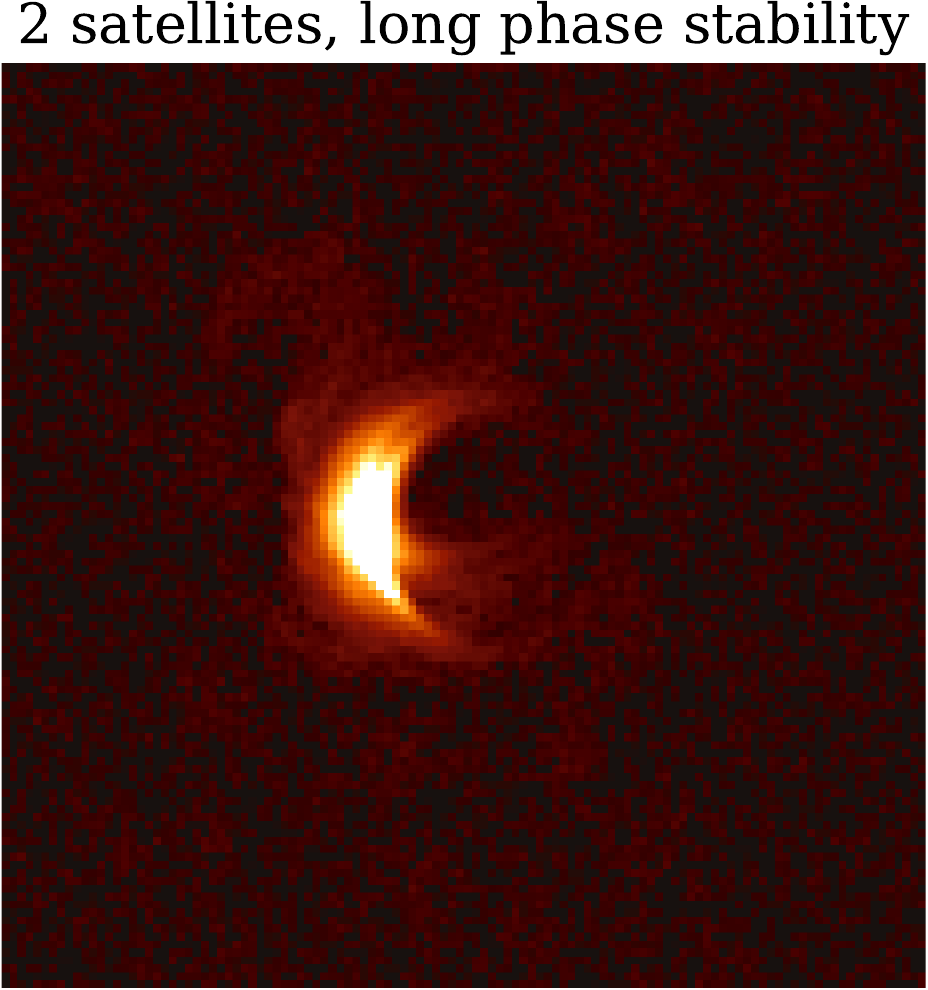}
\includegraphics[scale=0.36]{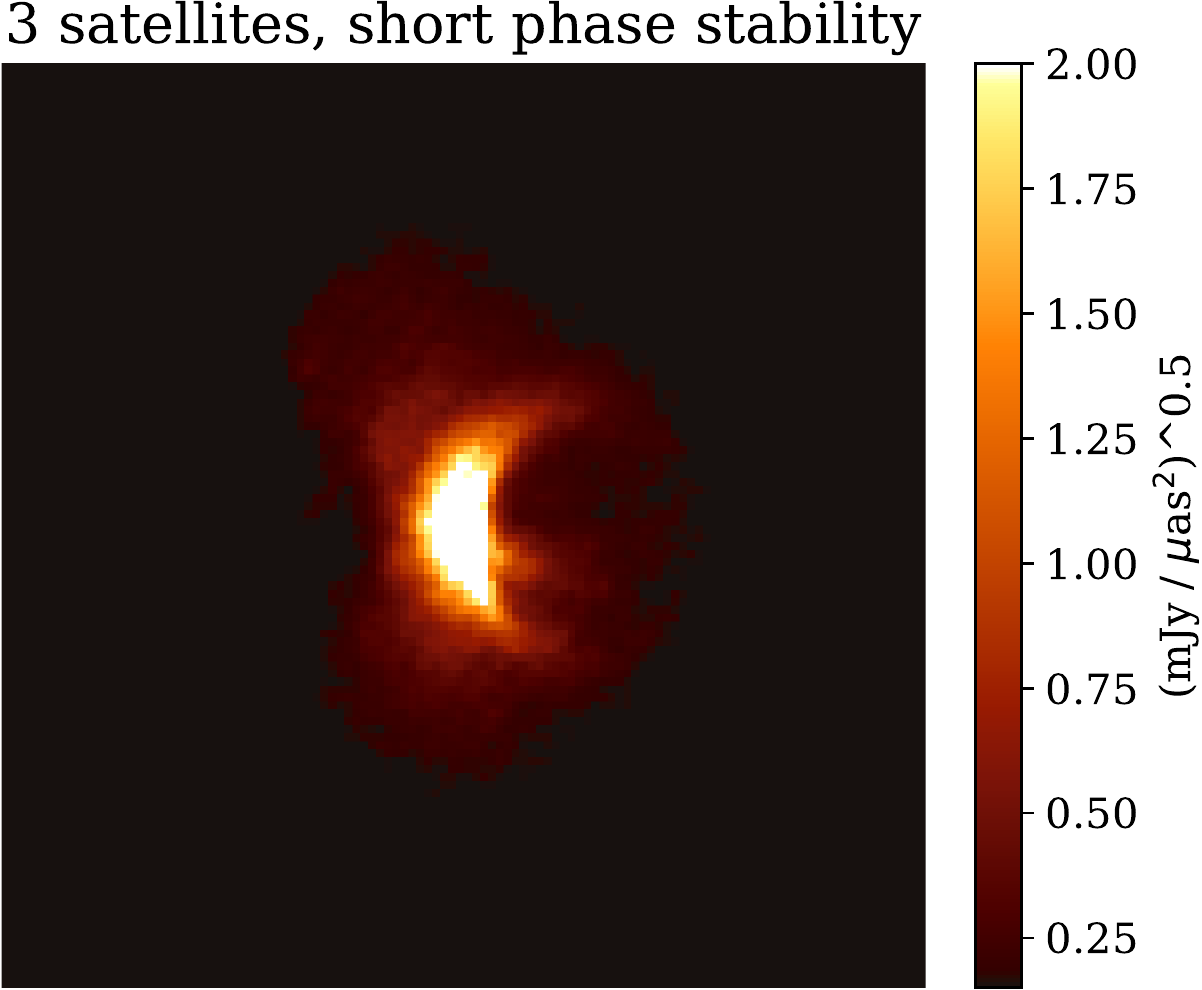}
\caption{\textit{Left:} Time-averaged GRMHD model of Sgr~A* \citep{Moscibrodzka2014}; Middle and \textit{Right:} Image reconstructions with the EHI consisting of two satellites with long phase stability allowing for the use of complex visibilities for imaging (Middle), and with the EHI consisting of three satellites with short phase stability relying only on the bispectrum for imaging (Right). The total integration time is 6 months for both cases. EHI reconstructions from \citet{RoelofsFalckeBrinkerink2019a}.}
\label{fig:ehisims}
\end{figure}

\citet{RoelofsFalckeBrinkerink2019a} performed imaging simulations of the EHI concept. Using GRMHD models of Sgr~A* at 690 GHz from \citet{Moscibrodzka2014} as input, complex visibilities were sampled at the EHI $uv$-spiral points, and thermal noise was added based on preliminary system parameters. Orbit radii of 13,892 km and 13,913 km were assumed for the two satellites, which gives a nominal resolution of 3.6 $\mu$as at 690 GHz. Each satellite was assumed to carry an antenna with a diameter of 4.4 m, which would fit in an ESA Ariane~6 launcher. Of course, larger dishes would yield even better results. Fig.~\ref{fig:ehisims} shows a GRMHD model image and its reconstructions for two EHI system variants. The middle panel shows the case of a perfectly phase stable EHI configuration with two satellites, limited by thermal noise only. Because of the dense and isotropic $uv$-coverage, the visibilities could be gridded in the $uv$-plane and an image could be reconstructed by taking the FFT of the complex visibilities, which were averaged over six iterations of the $uv$-spiral to build up signal-to-noise ratio on long baselines. The reconstructed image shows many of the detailed features that are present in the input GRMHD model. \citet{RoelofsFalckeBrinkerink2019a} show that visibility averaging also helps mitigating source variability (which occurs on a timescale of minutes for Sgr~A*), allowing to reconstruct an average image of a variable source. The basic reason is that averaging in Fourier space is the same as averaging in image space and the average structure is dominated by GR, which is not changing.

In practice, the EHI may not be phase stable over multiple months, depending on the attainable post-processing orbit reconstruction accuracy and clock stability. If the system is phase stable within an integration time (which is limited to timescales of minutes because of visibility smearing on arcs swept out in the $uv$-plane), a system consisting of three satellites would allow for the use of closure phase, which is the sum of the complex visibility phases on a triangle of baselines. Closure phases are robust against station-based phase errors such as those resulting from an inaccurate orbit model. The right panel of Fig.~\ref{fig:ehisims} shows a reconstructed image for such a system, made with the maximum entropy method implemented in the EHT-imaging library \citep{Chael2016,Chael2018}. The image quality is slightly less than for the idealised phase stable system, but the model features and size and shape of the black hole shadow are still recovered robustly. 

\section{Mission outlook and key technologies}

The THEZA concept aims at addressing multi-disciplinary cutting edge topics of modern astrophysics. While many of these topics presented in Section~\ref{Section_Science} are complementary in their science contents and synergistic in terms of engineering implementation requirements, a single mission addressing them all would likely be in the L-class category.  However, optimisation of the THEZA mission science composition might lead to lowering some technical requirements (e.g., frequency band coverage, data acquisition rate, number of space-borne elements, etc.) and therefore shifting the mission toward the M-class envelope. We also note a significant overlap in technology requirements and, partially, in science rationale of the THEZA concept with the concepts of the Origins Space Telescope\footnote{https://asd.gsfc.nasa.gov/firs/, Origins Space Telescope site, accessed 2020.12.27} and a mission for the far-infrared spectral domain with sub-arcsecond angular resolution, submitted as White Papers for the current ESA Voyage 2050 Call for Proposals, \citet{2020arXiv201202731W} and \citet{2020arXiv200206693L}, respectively. 

In general, all key engineering components required for THEZA implementation are well within the mainstream developments of relevant Earth-based and space-borne technologies. While we foresee a detailed analysis of the TRL figures for THEZA implementation at the stage of pre-design study of a specific mission, a preliminary evaluation conducted by the EHI study team has not identified insurmountable technological problems preventing a project with the launch well within the Voyage 2050 time frame.

VLBI is international and in fact global in its very nature. Obviously, this is even more so for Space VLBI. Not surprisingly, all three implemented SVLBI experiments and missions to date, as described in subsection \ref{Subsec_SVLBI}, have been widely international. We foresee that the THEZA concept implementation in a form of a specific mission would benefit greatly if it involves more than one major space agency. Such a collaboration not only enhances the mission potential by choosing the best available technologies but also might help in fulfilling budgetary limitations of all parties involved. The THEZA Core Team is aware of several highly compatible initiatives prepared within the ongoing US Decadal Astronomy Survey. Close coordination and collaboration with respective projects in the US, as well as in other countries, will be highly beneficial for THEZA implementation.

This White Paper presents the concept developed within the framework of ongoing studies in Europe, the United States, Japan, and Russia. A series of special workshops was initiated in 2018 with the first workshop held in Noordwijk, the Netherlands, in September 2018 (a part of its materials published and referred to in \citet{JASR2019}), and the second one in Charlottesville, VA, USA, in January 2020 \citep{2020arXiv200512767L}, to discuss scientific and technological issues of what is called here THEZA. We expect that this series of workshops will provide important contribution toward advancing the THEZA concept.

\begin{acknowledgements}
We are grateful to Beata~Budai for assistance in preparation of the artwork used in this paper and an anonymous referee for very useful comments and corrections.  

\end{acknowledgements}


\bibliography{bibliography}   % name your BibTeX data base

\end{document}